\documentclass[pra, twocolumn, preprintnumbers, superscriptaddress, eprintnumbers]{revtex4-1}
\usepackage[]{graphicx}
\usepackage{epstopdf}
\usepackage{bm}
\usepackage{framed}
\newcommand{\Mn}{Mn$_{50}$Ni$_{40}$Sn$_{10}$}

\usepackage{wasysym} 
\usepackage{hyperref}

\begin{document}
\title{Suppression of spinodal instability by disorder in an athermal system}
\author{Tapas Bar}\email{tb15rs042@iiserkol.ac.in}
\affiliation{Indian Institute of Science Education and Research Kolkata, Mohanpur, Nadia, West Bengal 741246, India}
\author{Arup Ghosh}
\affiliation{Indian Institute of Science Education and Research Pune, Pune 411008, Maharashtra, India}
\author{Anurag Banerjee}
\affiliation{Institut de Physique Th\'{e}orique, Universit\'{e} Paris-Saclay, CEA, CNRS,F-91191 Gif-sur-Yvette, France}
\preprint{Physical Review B {\bf 104}, 144102 (2021)}
\begin{abstract}
We observed asymmetric critical slowing down and asymmetric dynamical scaling exponent in the superheating and supercooling kinetic processes during the thermally-induced metal-insulator transition of MnNiSn based heusler alloy. During the transition to the insulator phase, the critical-like features get enhanced compared to the transition back to the metal phase. These experimental findings suggest that the metastable phase in the cooling branch of hysteresis has approached close to the spinodal instability. On the other hand, the extended disorder, generated over and above the intrinsic crystal defects during heating, triggers the excess heterogeneous nucleation before reaching the spinodal point. Zero temperature random field Ising model (ZTRFIM) simulation, inscribed for the athermal martensitic transitions, support the argument that the disorder smears the spinodal instabilities as the correlation length is bounded by the average distance between the disorder points.
\end{abstract}
\maketitle

\section{Introduction}
Hysteretic transition is an exciting subclass of abrupt thermodynamic phase transition (ATPT) where equilibrium thermodynamics breaks down on account of the system has accessed the metastable phase \cite{binder_rpp, Debenedetti, Planes_book}. Thermal hysteresis implies that there is a  discontinuous jump from one metastable minimum to another free energy minimum and often accompanied by rate-dependent effect \cite{Liu_prl16, Planes_prl04}, exchange bias \cite{exchange_bias}, kinetic arrest \cite{kinetic_arrest}. Such non-ergodic behavior is generally believed to arise from an interplay of disorder, thermal fluctuations, and activation barriers separating the two phases \cite{Planes_book}. When thermal fluctuation is insignificant in the kinetics of phase transformation (athermal), the metastable phase of a system can persist right up to spinodals (mean-field concept) where the activation barrier against nucleation vanishes \cite{Binder_pra84}. One would expect divergence of correlation length along with divergence of the relaxation time scale of order parameter due to the diffusive nature of dynamics \cite{Binder_pra84}.  This kind of athermal transition arises due to the suppression of fluctuation by the long-range force during the magnetic and structural phase transition of many complex functional materials, including manganites, transition metal oxides, etc. \cite{Planes_prl01, Zacharias_prl12, Rasmussen_prl01, Bar_prl18, Procaccia_pnas17}. Near spinodal, the ramified nucleating droplet diverges in all directions, unlike classical nucleation, where a single droplet of stable phase grows in a compact form \cite{klein-monette}. When the system approaches spinodal, the nucleation rate becomes very slow,  indicate spinodal slowing down \cite{Kundu_prl20}. The nucleation rate decrease further as the range of interaction increases and finally goes to zero when the range of interaction becomes infinity \cite{klein-monette}, then the transition becomes mean-field like \cite{Binder_pra84, Klein_pseudospinodal}. The recent experiment on dynamic hysteresis scaling supported by numerical analysis suggested the mean-field like spinodal instability exists in the correlated system \cite{Bar_prl18, Basov_nat18}.

However, the disorder is known to yield heterogeneous nucleation of athermal martensitic transformation \cite{Cao_prb90} characterized by jerky propagation related to avalanches \cite{Planes_prl01, Jerky_DTA_peak}. The disorder may reduce the free energy barrier of nucleation before arriving at the spinodal  \cite{Debenedetti, Klein_pre07}. The influence of disorder on the spinodal has not yet been fully explored except some model system \cite{Nandi_prl16, Zapperi_Roy, Liu_prl16, Sethna_prl93, Zheng_prb02}. Although disorder associated athermal transition is found in many complex functional materials undergoing hysteretic transition \cite{Planes_book}, glassy materials \cite{Ozawa_PNAS18}, earthquake \cite{earthquakes}, QCD patterns formation \cite{QCD_prd20}, social and economic systems \cite{social}.

This article focuses on one such system where a kinetic asymmetry in the supersaturated transition arises from the extended disorder generated during the superheating process. We experimentally observed that, even at finite temperature, the transition is independent of thermal fluctuation, i.e., athermal. One would expect to see the footprints of diverging susceptibility in its temporal features as the fluctuation-less metastable phase can survive up to the unstable singularity (spinodal) under phase kinetics \cite{Procaccia_pnas17, Bar_prl18, Planes_prl01, Zacharias_prl12}. However, extended disorder reduces the degree of superheating through nucleation on the kinetic-path before vanishing the activation barrier. In this letter, through the dynamic hysteresis and critical slowing down measurements supported by ZTRFIM simulations, we first experimentally report that in an athermal system, the extended disorder overrules the spinodal instabilities via heterogeneous nucleation.

\section{Experimental Results}

\begin{figure}[!h]
\center
\includegraphics[scale=0.32]{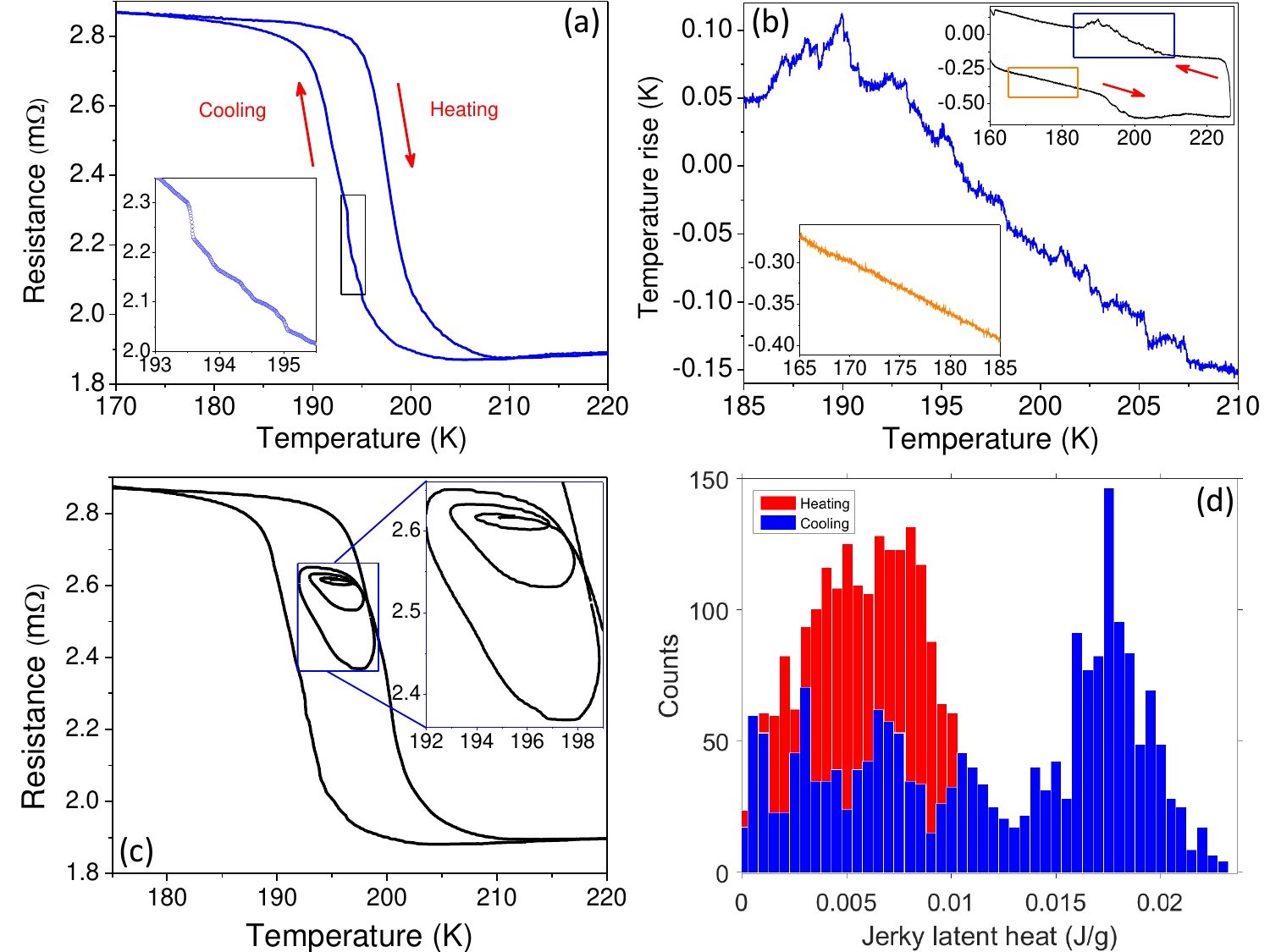}
\caption{(a) The avalanche-like jump in resistance during the transition. (b) Jerky like peak during the transition seen in DTA measurement (right inset). The off-transition DTA signal is shown for comparison (left inset). (c) Return point memory for three partial loops. The inset shows returning points more clearly in sub-subloops. (d) Histogram of jerky latent heat as a function of magnitude. The heat fluxes were measured in 1/7 sec interval, and the corresponding temperature scanning rate was 6 K/min.}
\label{Avalanche}
\center
\end{figure} 

Figure \ref{Avalanche} (a) shows that a typical resistance measurement done as a function of temperature during the first-order phase transformation (FOPT) at 200 K in polycrystalline heusler alloy. The details of sample preparation and characterization are given in Appendix \ref{App:Sample}. There is a thermal hysteresis (width $\sim$ 5 K) associated with the change in resistance of austenite transition ($\sim$ 197 K) during heating and the martensitic transition ($\sim$ 192 K) during cooling. The electronic transition, coupled with a magnetic and a structural transition, occurs through a series of avalanches. Each avalanche is accompanying with a jump in resistance [Fig. \ref{Avalanche} (a) (inset)] and a corresponding jerky latent heat [Fig. \ref{Avalanche}(b)] \cite{Jerky_DTA_peak}. The latent heat of transitions is evaluated from the differential thermal analysis (DTA) signal [Fig. \ref{Avalanche}(b) right inset] by numerically fitting an equivalent model for the setup. A more detailed description of the home-made DTA experimental setup can be found elsewhere \cite{Bar_rsi21}. Figure \ref{Avalanche} (d) shows the distribution of jerky heat as a function of their size. The jumps during cooling are larger than the heating implies that the disorder linked with the austenite transition is more compare to the martensitic transition \cite{Sethna_prl93, Sethna_prl95}. Moreover, these asymmetric jumps are independent of the driving rate (for details, refer to Appendix \ref{App:Avalanches}). Such kinetic asymmetry emerges due to the heterogeneous nucleations at two distinct disorder points: extended disorder \cite{Disorder_comment} (structural twin walls \cite{Jerky_DTA_peak,Fan_prb11,Cao_prb90}, surface-localized defects \cite{Kang_ass14}, dislocation \cite{Perez-Reche_prl07}, etc) which appear during heating branch of hysteresis on top of intrinsic disorder which remains in both branches \cite{Fan_prb11, Disorder_comment}. Return point memory of hysteresis loop [Fig. \ref{Avalanche} (c)] implies the existence of intense disorder (above the critical value $\sigma_c$) associated with ATPT \cite{Sethna_prl93,Disorder_RPM}.

\begin{figure}[!h]
\center
\includegraphics[scale=0.32]{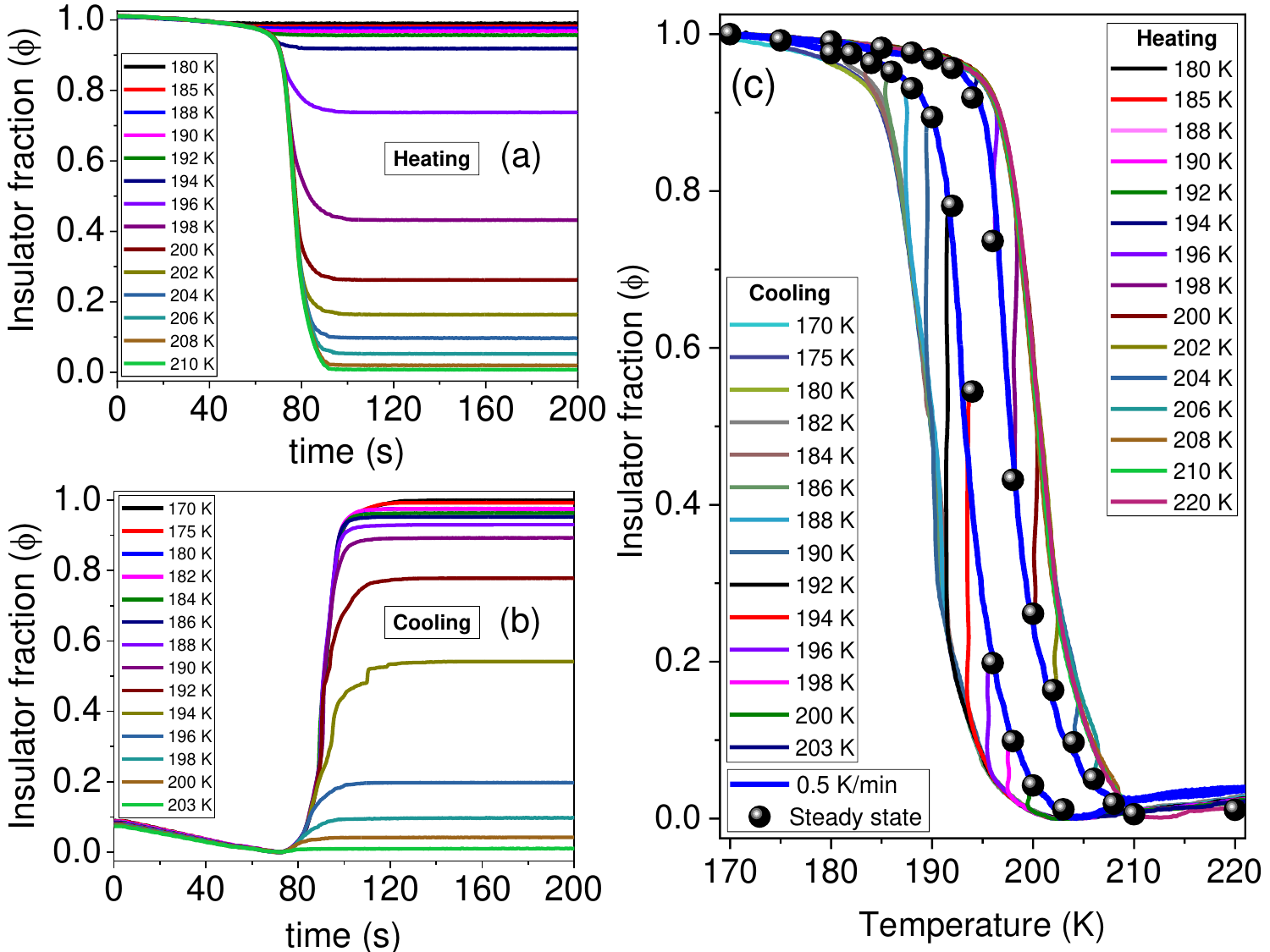}
\caption{Order parameter as a function of time (a),(b), and temperature (c) acquired from quench-and-wait experiments for different waiting temperatures. The order parameter finally reaches the steady-state value (\textbullet\ within c) for a given sufficient wait. The steady-state values coming out from dissipative phase ordering experiments are very close to the dynamic hysteresis measurement of ramp rate 0.5 K/min (blue-line inside c). The temperature quench rate of the measurements was 40 K/min.}
\label{quench_wait}
\center
\end{figure}

Thermal hysteresis implies discontinuity in free energy and discontinuity in entropy at the transition points. Therefore, this transition can be termed as `zeroth-order' \cite{Bar_prl18, Catastrophe_theory}.  The continuation of free energy beyond the expected transition point (where the free energy of two phases intersect) enlarge the lifetime of metastable state \cite{binder_rpp, Catastrophe_theory}. Consequently,  the metastable state act like an equilibrium state and the transition becomes athermal. Such continuation breaks at the limit of stability due to the essential singularity of spinodal points. At these points, the relaxation time of the metastable state is expected to diverge \cite{Analytic_continuation}. The detailed mean-field free energy formalism of thermally driven hysteretic transition has been discussed in Appendix \ref{App:Free_energy}.

However, the phase ordering relaxation time is measured by the quench and hold experimental technique. Starting from an initial temperature of 120K (or 250K), which is far from transition regions, the sample is heated up (or cool down) rapidly to a specific set temperature ($T_R$), then allowed to equilibrate for sufficient time. The temporal evolution of resistance is represented in terms of insulator fraction (order parameter) \cite{Insulator_frac}. The detailed calculations of insulator fraction are given in Appendix \ref{App:QAW}. Figure \ref{quench_wait} shows the insulator fraction for different set temperatures for heating and cooling quenches. After an adequate wait time ($\sim$ 60 s), the system reaches the quasi-static value. After that, no temporal evolution of order parameter is observed (See Appendix \ref{App:QAW}); suggest that the system, driven by thermal fluctuations, unable to jump the activation barrier and remains in the metastable local minima. Such behavior indicates an athermal transition \cite{Planes_prl01, athermal}, where the thermal fluctuations are suppressed by long-ranged strain-strain force \cite{Rasmussen_prl01} and the kinetics of transition is only control by external parameters. When the range of interaction is very large (Coulomb type \cite{Basov_Nat17}), the fluctuation-related barrier crossing becomes disabled, and the transition dynamic should follow the deterministic track—eventually, the transition acts like the mean-field one \cite{Bar_prl18}. Spinodal slowing down has been observed under deep supersaturation in the athermal system \cite{klein-monette, Klein_pseudospinodal, Bagchi_prl07}. This slowing down can manifest as the much-discussed delay in the onset of switching around the bifurcation points of the FOPT (i.e., finite time effects); the change in the hysteresis loop area $A(R)$  (or shift in transition points) with $R$, the rate of change in the field $H$ or temperature $T$ can dynamically scale as:
\begin{equation}\label{eq:1}
A(R)=A_0 + a R^\Upsilon.
\end{equation}
The quasi-static hysteresis loop area $A_0$ must be nonzero for such transitions \cite{Bar_prl18, Zheng_prb02, Rao, Fan_Zhong, Thermal_hysteresis, glycerol, Shukla_pre18, Pal_prb20}.  While the values of $\Upsilon$ have been obtained analytically \cite{Rao}, numerically \cite{Zheng_prb02, Fan_Zhong, Shukla_pre18}, and experimentally \cite{Thermal_hysteresis, glycerol, Bar_prl18} in different systems, for the field as well as temperature-driven FOPTs.  The mean-field exponent $\Upsilon = 2/3$ is believed to be the signature of spinodal like transition \cite{Bar_prl18, Fan_Zhong}.  The dynamical scaling that resembles a spinodal like transition occurs when an inherent slowing down leads to overshooting in the transition. Note that the power-law scaling of hysteresis is considered a scale-free behavior of the characteristic response time.

\begin{figure}[!h]
\center
\includegraphics[scale=0.32]{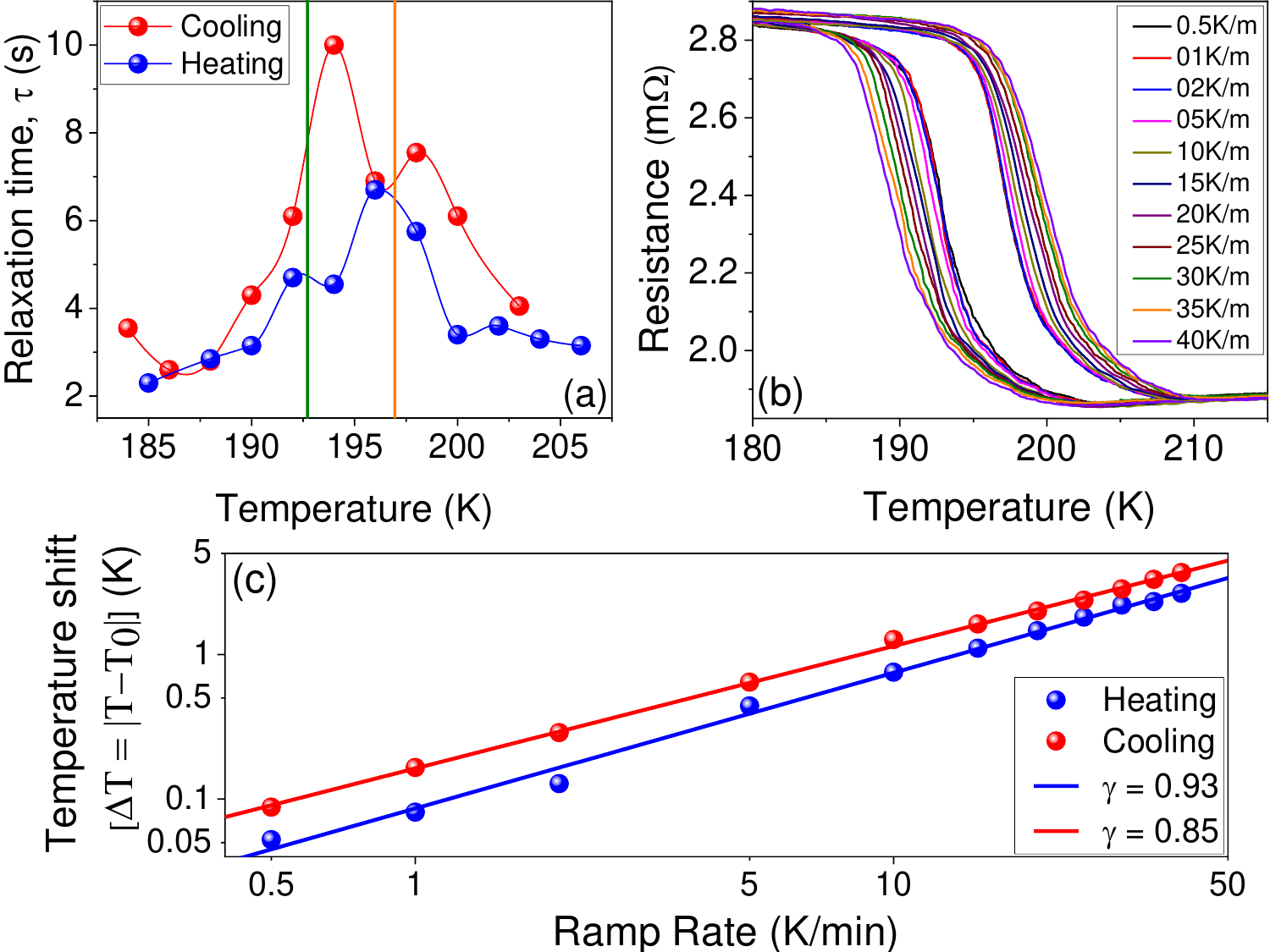}
\caption{(a) Temperature dependent phase ordering relaxation time constant. The time constant has been extracted by exponential curve fitting in the temporal evolution of order parameter of quench and hold experiments [Fig. \ref{quench_wait}]. The vertical lines corresponds to the quasi-static transition temperatures for cooling (green) and heating (orange). (b) Temperature dependent resistance for different ramp rates. (c) log-log plot of shift in transition temperatures with temperature scanning rates. The power laws fitting exponent coming out to be $\Upsilon \approx 0.93$ for heating and $\Upsilon \approx 0.85$ for cooling.}
\label{time_const}
\center
\end{figure}

However, the signature of spinodal slowing down reflects both in quench-and-hold and dynamic hysteresis measurements [Fig. \ref{time_const}]. The phase ordering relaxation time constant peaks at the transition points, but there is a clear qualitative difference between the heating and cooling branch of hysteresis [Fig. \ref{time_const}(a)]. The peaks are significantly milder in the heating branch, hints non-diverging time scale of relaxation. Such non-robust critical behavior in the athermal system arises from local fluctuations induced by the impurities, which acts as a heterogeneous nucleating site \cite{Imry_prb79}. The extended disorder, along with crystal impurities, is present during the transition to the metal phase. The physics is controlled by its disorder-induced fluctuations leading to a large rounding of the time scale divergence.  On the other hand, the transition back to the insulator phase has no extended disorder, and a comparatively sharper peak is observed. Besides, we also observe the asymmetric finite time effect of the transitions, the shift in transition temperature with the ramp rates R [Fig. \ref{time_const}(b)]. Although the dynamical renormalized shifts $\Delta T$ both for heating and cooling fulfil a scaling relationship $\Delta T (R)=|T^{i}_0-T^{i}_{obs}(R)|\propto R^\Upsilon$, $i=heat$ or $cool$ around two decades of ramp rates [\ref{time_const} (c)], the exponent are found to be dissimilar; $\Upsilon = 0.93 \pm 0.13$ for heating and  $\Upsilon = 0.85 \pm 0.07$ for cooling. The error in exponents has been calculated in Appendix \ref{App:Error}. The exponents are inconsistent with the mean-field spinodal value ($\Upsilon = 2/3$). Such non-mean-field behavior appears due to the coexistence of barrier-free spinodal nucleation under deep supersaturation \cite{Klein_pre07, klein-monette}, and barrier crossing heterogeneous nucleation at the impurity center \cite{Klein_pre07, Cao_prb90} which is more during heating compared to cooling. $\Upsilon > 2/3$ has so far only been seen in the experiment for glass transition of glycerol \cite{glycerol} and in (Monte-Carlo) simulation for Ising model \cite{Shukla_pre18} at the critical point.

\section{Simulation Results}

To confirm the above interpretation (disorder suppressed the spinodal instability), we have performed three-dimensional (3D) zero temperature (athermal) random field Ising model simulations in the context of quench-and-hold and dynamic hysteresis measurements where the magnetization (order parameter) is the response function of an external magnetic field (details can be found in Appendix \ref{App:ZTRFIM}). The nonequilibrium bistable system involving time-dependent bifurcation parameters such as magnetic field, temperature, laser-field (in bistable semiconductor laser) can be reduced to the spin system \cite{Zhong_pre95, RRoy_prl95, Thermal_hysteresis} as the instability point for such systems are similar to the fixed point of $\phi^3$ theory for field-driven transitions \cite{Fan_Zhong}. Therefore, even though the experimental system is temperature driven instead of field-driven, the ZTRFIM captures the essential nature of athermal transition, including random disorder \cite{Sethna_prl93, short_range}. The Hamiltonian of the model given by
\begin{equation}\label{eq:2}
\mathcal{H} = -J\sum\limits_{\langle i,j\rangle} s_is_j - \sum\limits_{i} [H(t)+h_i]s_i \ ,
\end{equation}
where Ising spins $s_i = \pm 1$, placed on the 3D lattice, coupled with nearest-neighbor pairs through coupling strength $J$. $H(t)$ is the external field, and $h_i$ represents the random field disorder, taken from a Gaussian distribution of zero mean and variance $\sigma$. $\sigma$ represents the disorder strength.

\begin{figure}[!h]
\center
\includegraphics[scale=0.29]{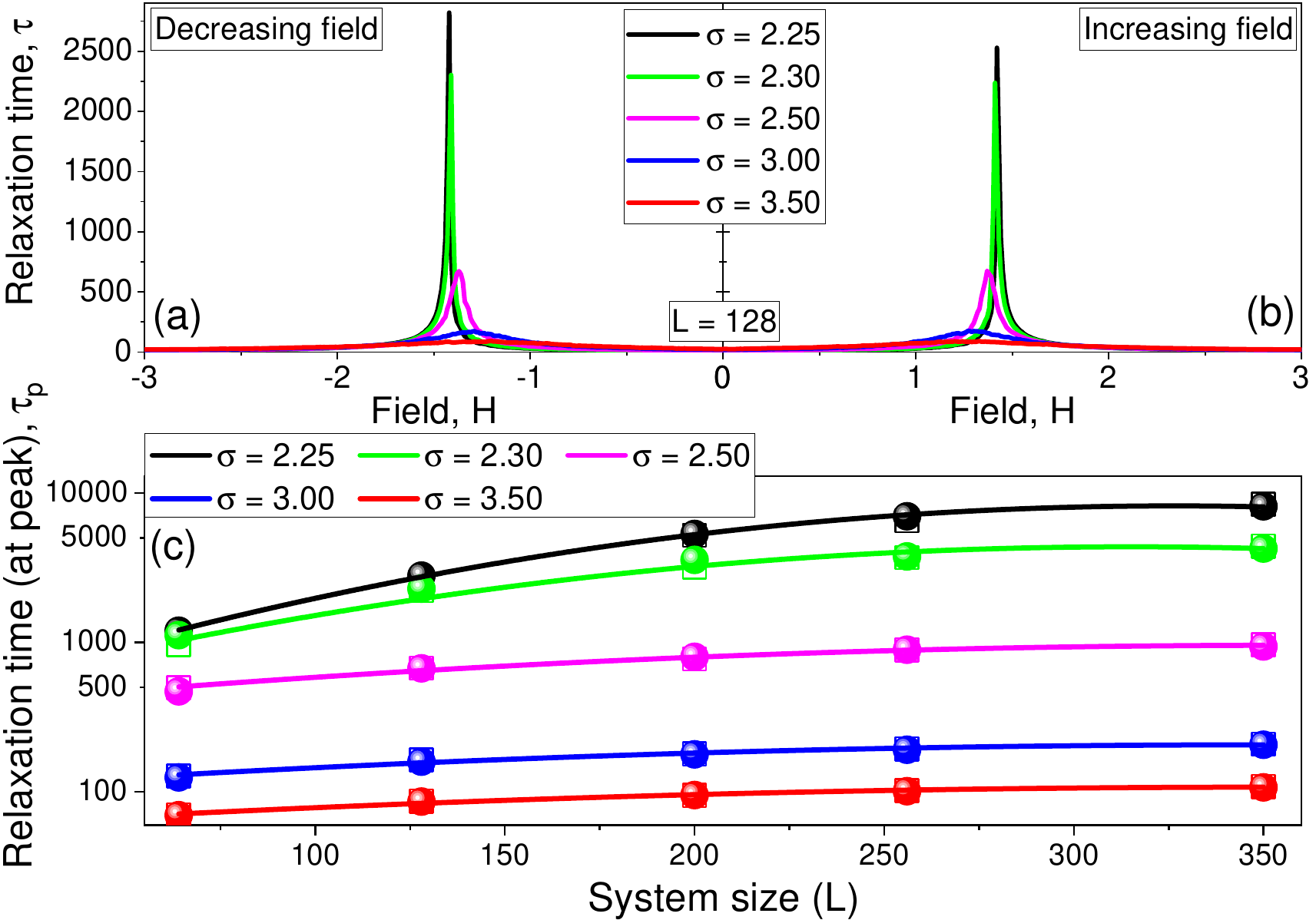}
\caption{Phase ordering relaxation time constant during wait after rapid increasing (a) and decreasing (b) fields drawn from ZTRFIM simulations of size $128^3$ for disorder strength $\sigma$ = 2.25, 2.3, 2.5, 3, and 3.5. (c) The time constant peak values for increasing (\textbullet) and decreasing (\Square) fields as a function of system size for different disorder strengths. Note that a little mismatch of peak height during field increasing and decreasing for a particular disorder strength depends upon how close one can reach the transition points.}
\label{eq_time}
\center
\end{figure}

The quench-and-hold experiments were performed in the ZTRFIM system on a cubic lattice of size $L^3$ under periodic boundary conditions with disorder $\sigma > \sigma_c$. The $\sigma_c$, critical disorder, separates a high-disorder regime where infinite avalanche never happens, i.e., the field-versus-magnetization hysteresis loop is continuous and a low-disorder regime where hysteresis loop contains macroscopic jump \cite{Sethna_prl93, Nandi_prl16}. The value of $\sigma_c = 2.16$ and $2.23$ for $320^3$ and $60^3$ system respectively \cite{Sethna_prl95,Sethna_prl93}. Although $\sigma_c$ is now known up to the five decimal point of accuracy [$\sigma_c(L \to \infty)=2.27205$] for the infinite system size limit \cite{Fytas_prl13,Fytas_pre16}. We choose a fully polarized initial state, and then the magnetic field is abruptly set to a specific value close to the transition point as the system is allowed to equilibrate at $T = 0$. The thorough numerical protocols are discussed in Appendix \ref{App:ZTRFIM_time}. The equilibration time (relaxation time) sharply peaks at the transition, although the peak height decreases with the increasing disorder strength [Fig. \ref{eq_time}].  The relaxation time peak grows slowly with system size, showing a mild finite-size effect, unlike power-law growth in the case of classical criticality followed by spinodal transition with critical disorder. This manifests the non-robust critical behavior of spinodal transition.

\begin{figure}[!h]
\center
\includegraphics[scale=0.3]{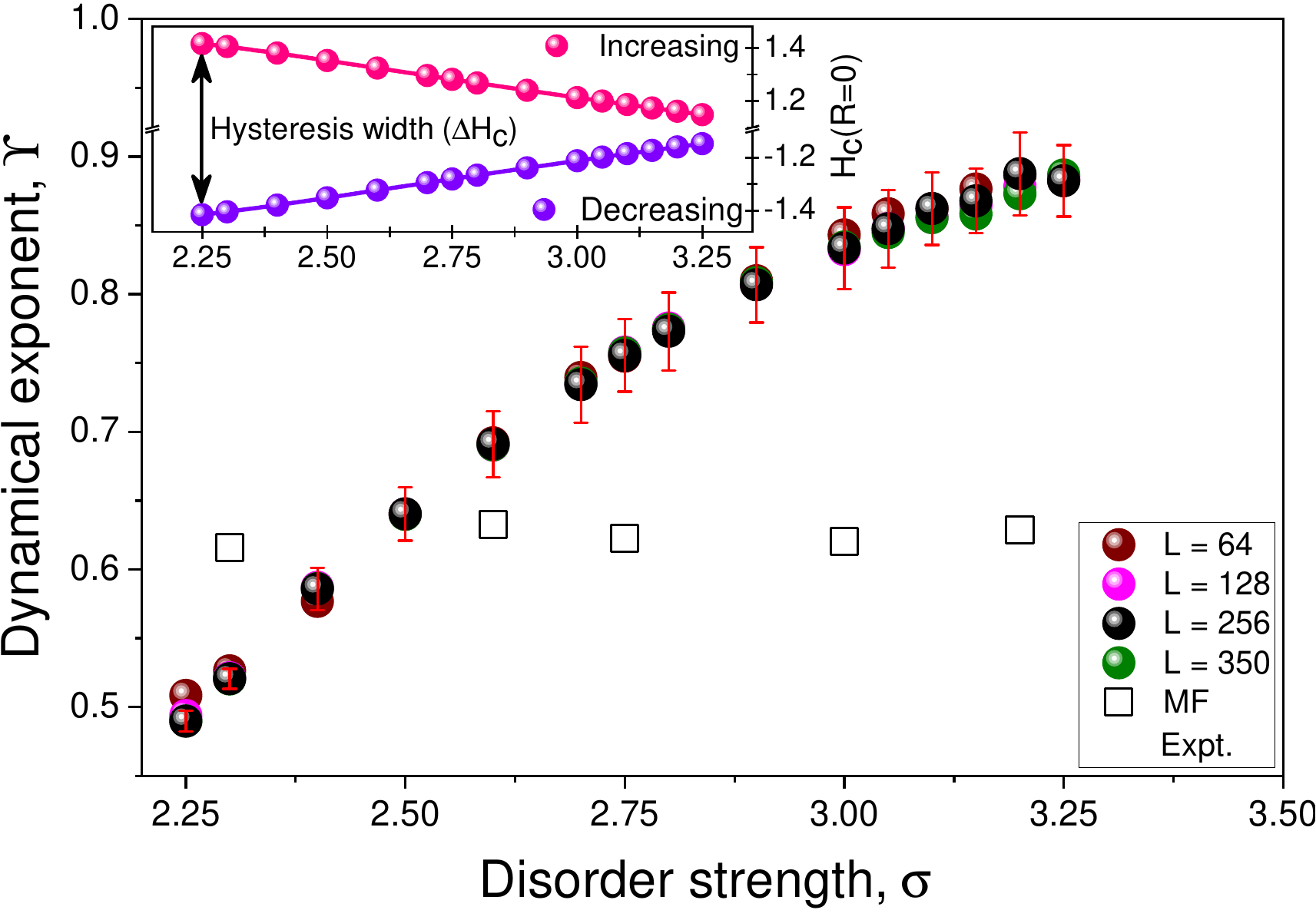}
\caption{Dynamical hysteresis scaling exponent ($\Upsilon$) versus disorder strength ($\sigma$) yield in the 3D-ZTRFIM simulations for different system size. ({\large $*$}) represent the experimentally obtained exponent values for cooling (blue) and heating (orange). (\Square) stand for zero temperature mean-field dynamical exponent, $\Upsilon \approx 0.62 \pm 0.01$. The steady-state hysteresis width decreases with increasing disorder for $\sigma > \sigma_c$ (inset).}
\label{Dynamic_hysteresis}
\center
\end{figure}

To quantify the role played by disorder on spinodal transition, we study the dynamic hysteresis in the ZTRFIM. The systematic shift in coercive fields $H_c(R)$, the fields at which the transition takes place, with the field rates R fulfill a scaling relationship similar to the experiment [Eq. \ref{eq:1}] where the steady-state coercive field $H_c(0)$ is no longer a free parameter (see Appendix \ref{App:ZTRFIM_rate} for details). Here no excess disorder (unlike experiments) is induced either during increasing or decreasing the field; consequently, the asymmetric exponents are not expected and observed. Most importantly, the exponent $\Upsilon$ increases with the increasing disorder, and the scaling does not form when the disorder crosses a threshold level, $\sigma_{th}$ = 3.30 \cite{Sethna_prl95} [Fig \ref{Dynamic_hysteresis}]. $\Upsilon$ is independent of system size as it is directly linked with the time scale divergence \cite{Footnote}. Above $\sigma_{th}$, disorder-induced nonperturbative athermal fluctuations destroy the spinodal singularity, and transition occurs through percolation \cite{Moreira_prl12}. The experimentally observed exponents emphasize that the cooling cycle is more critical than the heating.

\section{Conclusion}
Based on the RFIM calculations, we argue that for the systems with low disorder the transition is governed by a small number of relatively large macroscopic jumps. In contrast, systems with large disorder display smooth transitions with a large number of small-size avalanches, and the lopsided disorder gives rise to asymmetric avalanches in the superheating and supercooling kinetic processes. Heterogeneous nucleating droplets emerge from disorder-induced local fluctuation on the kinetic-path before reaching the spinodal instability. However, the thermal growth of the droplets suppressed by a long-range interacting force leads to a non-mean field spinodal slowing down: relaxation times are peaking but not diverging and show mixed order (first order and continuous) fracture at the transitions \cite{Shekhawat_prl13, Rizzo_prl13}. The crossover from the robust mean-field spinodal, i.e., with diverging correlation length and susceptibility, to the non-robust mixed transition, exhibiting nonuniversal dynamical exponents, takes place due to the finite correlation length whose growth is bounded by the average distance between the disorder points \cite{Imry_prb79}. As disorder increases, the heterogeneous nucleation increases, which leads to the suppression of spinodal more and more, and finally reaches a threshold level where a distinct crossover takes place from spinodal-like to percolation behavior \cite{Shekhawat_prl13}.

Finally, we conclude by saying that when the thermal fluctuation is irrelevant, disorder-induced local fluctuation smeared the spinodal instabilities, and our results established the existence of spinodal-singularity beyond Mean-field. Similar behavior has been seen in simulations of low-disorder RFIM \cite{Nandi_prl16}, 2D-Ising model \cite{Scheifele_pre13}, and Kob-Andersen model \cite{Bhowmik_pre19}.

\section{Acknowledgements and Author contributions}
It is a pleasure to thank Bhavtosh Bansal for providing access to the lab equipment and useful discussions. We also thank Prabodh Shukla, Varsha Banerjee, Fan Zhong for helpful comments and suggestions. A.B. acknowledges post-doctoral funding from ERC, under grant agreement AdG694651-CHAMPAGNE.
 
A.G. prepared the sample. A.B. carried out the simulations. T.B. performed the measurements, analyzed the data, conceived the problem, and wrote the paper.

\appendix
\section{Sample Preparation and characterization}
\label{App:Sample}
The \Mn\ off-stoichiometric Heusler alloy is prepared using an arc-melting technique inside a 4N purity argon atmosphere. The metals (Mn, Ni, and  Sn) are more than 99.9 \% pure. We seal the as-prepared ingot separately in evacuated quartz ampoules to anneal the samples at 900 $^{\circ}$C for 96 h. We immediately cool the ampoule by quenching it into the ice water. The crystallographic parent phase of the samples has been confirmed from the room temperature X-ray diffraction (XRD) patterns, recorded using CuK$_{\alpha}$ radiation. Finally, we perform the compositional analysis by energy dispersive analysis of X-ray attached with a field emission scanning electron microscope.

\begin{figure}[!b]
\center
\includegraphics[scale=0.3]{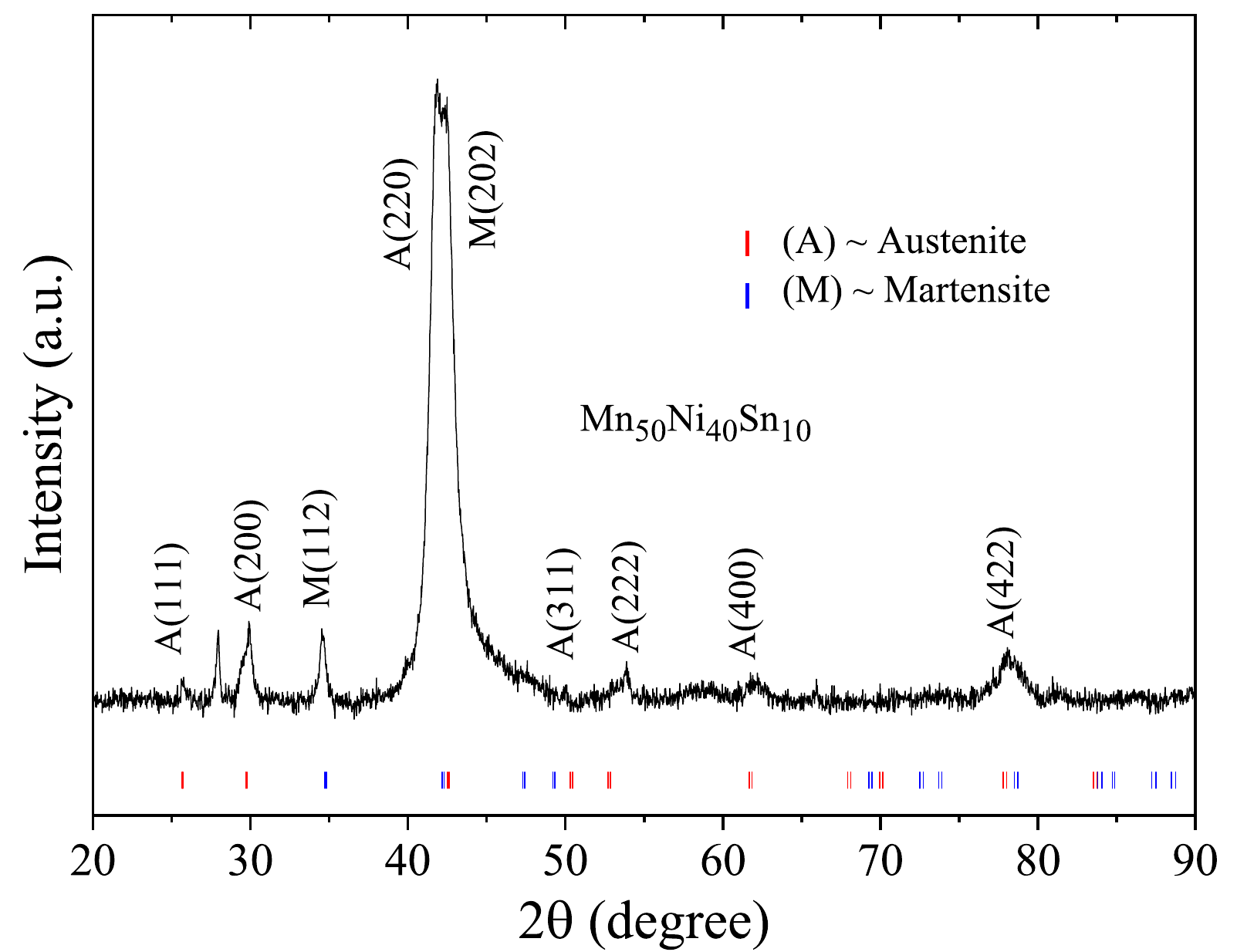}
\caption{Room temperature XRD of \Mn\ Heusler alloy using CuK$_{\alpha}$ radiation }
\label{XRD}
\center
\end{figure}

\begin{figure}[!h]
\center
\includegraphics[scale=0.3]{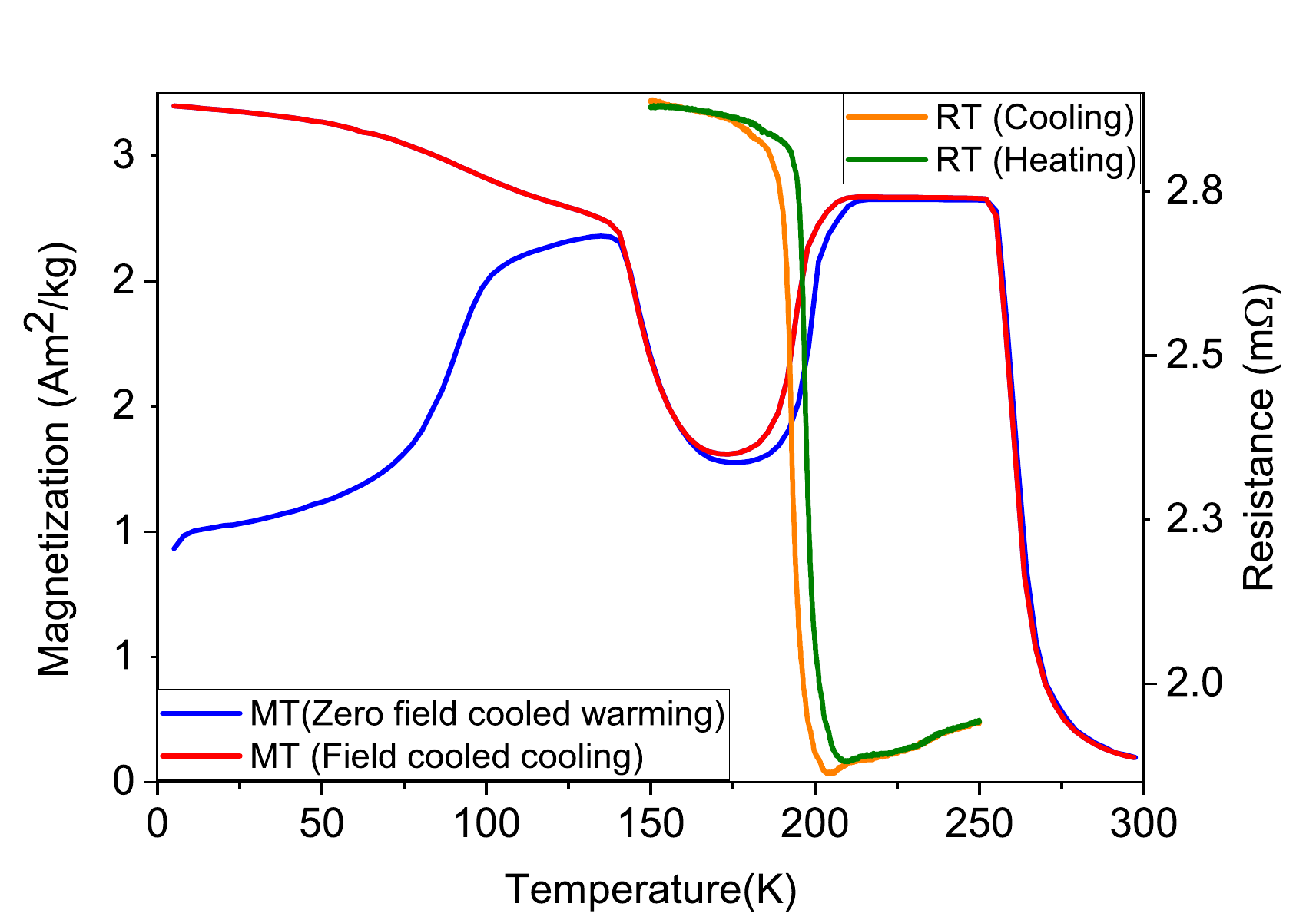}
\caption{Magnetization versus temperature and resistance versus temperature of a polycrystalline Mn-Ni-Sn based Heusler alloy.}
\label{MT_RT}
\center
\end{figure}

Structural and magnetic properties:
Figure \ref{XRD} shows the room temperature XRD pattern for off-stoichiometric \Mn\ Heusler alloy, which confirms the presence of cubic austenite (Hg$_2$CuTi- type) phase  \cite{XRD}. One can notice the co-existence of a small amount of tetragonal martensite phase also. The temperature-dependent field cooled cooling (FCC), and zero-field cooled (ZFC) magnetization of the material are recorded under 100 Oe fields [Fig. \ref{MT_RT}]. Figure \ref{MT_RT} shows a magnetic transformation near 270 K, the ferro-para transition at Curie temperature (TCA) in the austenite phase. Around T = 200 K, the magnetization of the sample rapidly decreases on cooling. The material transforms from a highly magnetic cubic austenite structure to a weakly magnetic tetragonal martensite structure. A thermal hysteresis associated with this transition has been observed both in magnetic and transport measurements. The transition is recognized as a first-order magneto-structural transition. Below this temperature, the magnetization again increases due to the presence of a Curie point in the martensite phase. A temperature below 125 K, a glassy phase may exist, which might show the effect of the Exchange Bias \cite{Ghosh_jap16}.

Recalling the XRD data, we have found that the martensite phase exists near room temperature. The magnetic measurements in Fig. \ref{MT_RT} also show that the structural transition resides near 200 K. It is unusual to observe the XRD peaks of a low-temperature structure at a temperature 100 degrees above the transition. It is only possible if the sample has a significant amount of disorder and strain \cite{Ghosh_apl14, Ravel_apl02}.

\section{Rate dependent avalanche}
\label{App:Avalanches}
In the DTA experiments corresponding to Fig. \ref{Avalanche}(b) and \ref{Avalanche}(d), we obtain jerky latent heat peaks related to the formation of new phase nucleated at the defect points. Moreover, the amount of jerky heat released during cooling is more compare to the heating. The DTA curve in Fig. \ref{Avalanche} corresponds to the temperature scanning at the rate of 6 K/min. In the athermal ATPT, the system has to be driven above some minimum temperature or field rate, below which no avalanches can be detected due to insignificant excitations in the metastable states \cite{Planes_prl04}. On the other hand, the transition becomes continuous rather than a series of avalanches above this driving rate, where the time scale of each avalanche is smaller than the data acquisition time \cite{Avalanches_VO2}. In between 0.5 K/min and 10 K/min temperature scanning rates, we were able to measure the MIT through a series of avalanches. The driving-rate dependence of the critical exponents of the avalanche distribution in first-order phase transitions had been observed previously \cite{Planes_prl04, Liu_prl16}. However, the asymmetry in the superheating and supercooling kinetic processes are independent of the sweep rate \cite{Schuller_prb18}.

\begin{figure}[!h]
\center
\includegraphics[scale=0.3]{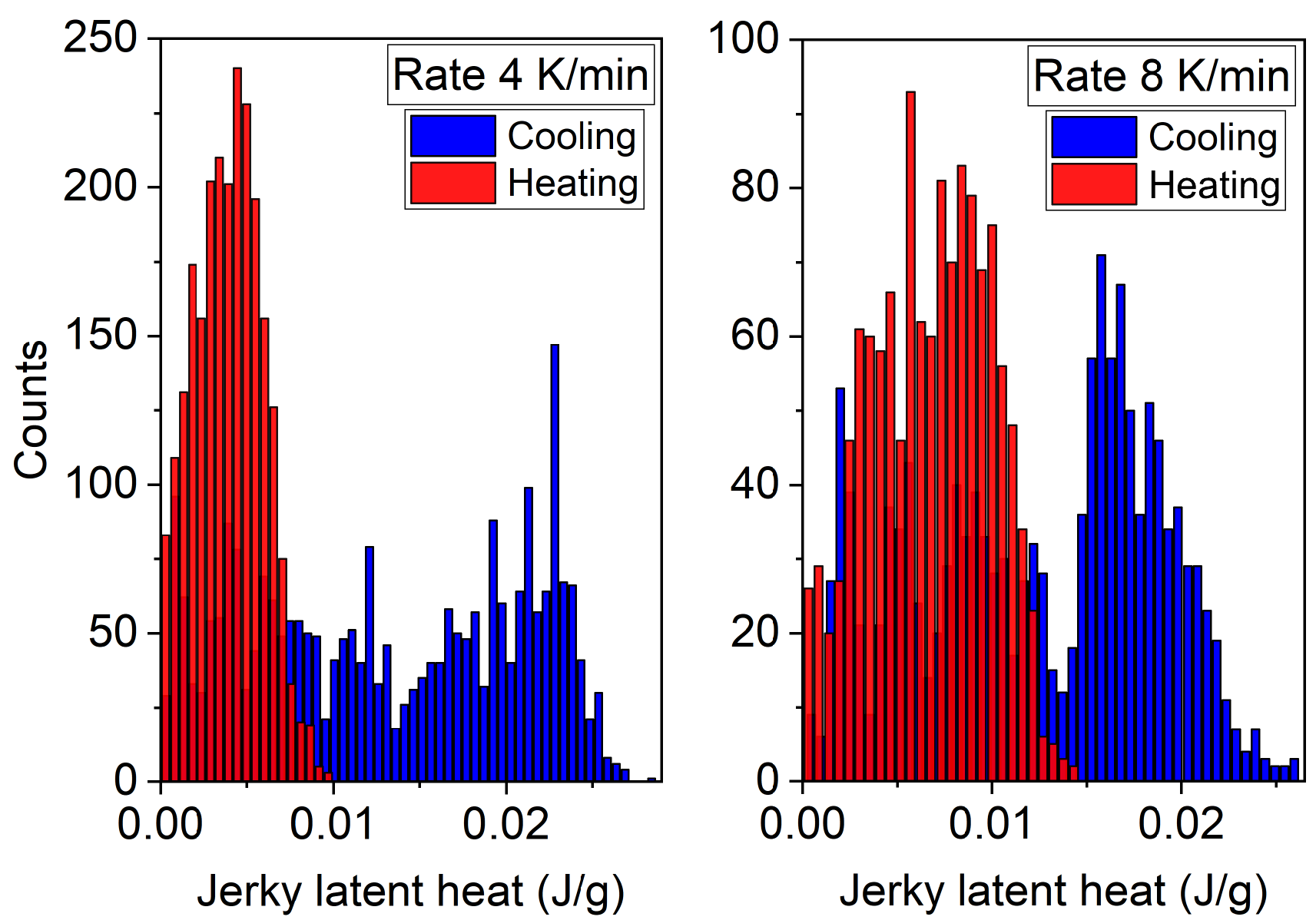}
\caption{Avalanche distribution of jerky latent heat as a function of magnitude, measured in 1/7 sec interval. The two curves correspond to two different temperature scanning rates: 4 K/min (left) and 8 K/min (right).}
\label{Avalanches_rate}
\center
\end{figure}

Figure \ref{Avalanches_rate} shows the latent heat jump distribution for two different temperature scanning rate of 4 K/min and 8 K/min. Both the curves manifest that the large jump appears in the cooling branch of the hysteresis. The driving rate-dependent results indicate that an apparent asymmetry is always observed for an individual rate within our measurement capabilities.

\section{Phase ordering time}
\label{App:QAW}
\begin{figure}[!b]
\center
\includegraphics[scale=0.3]{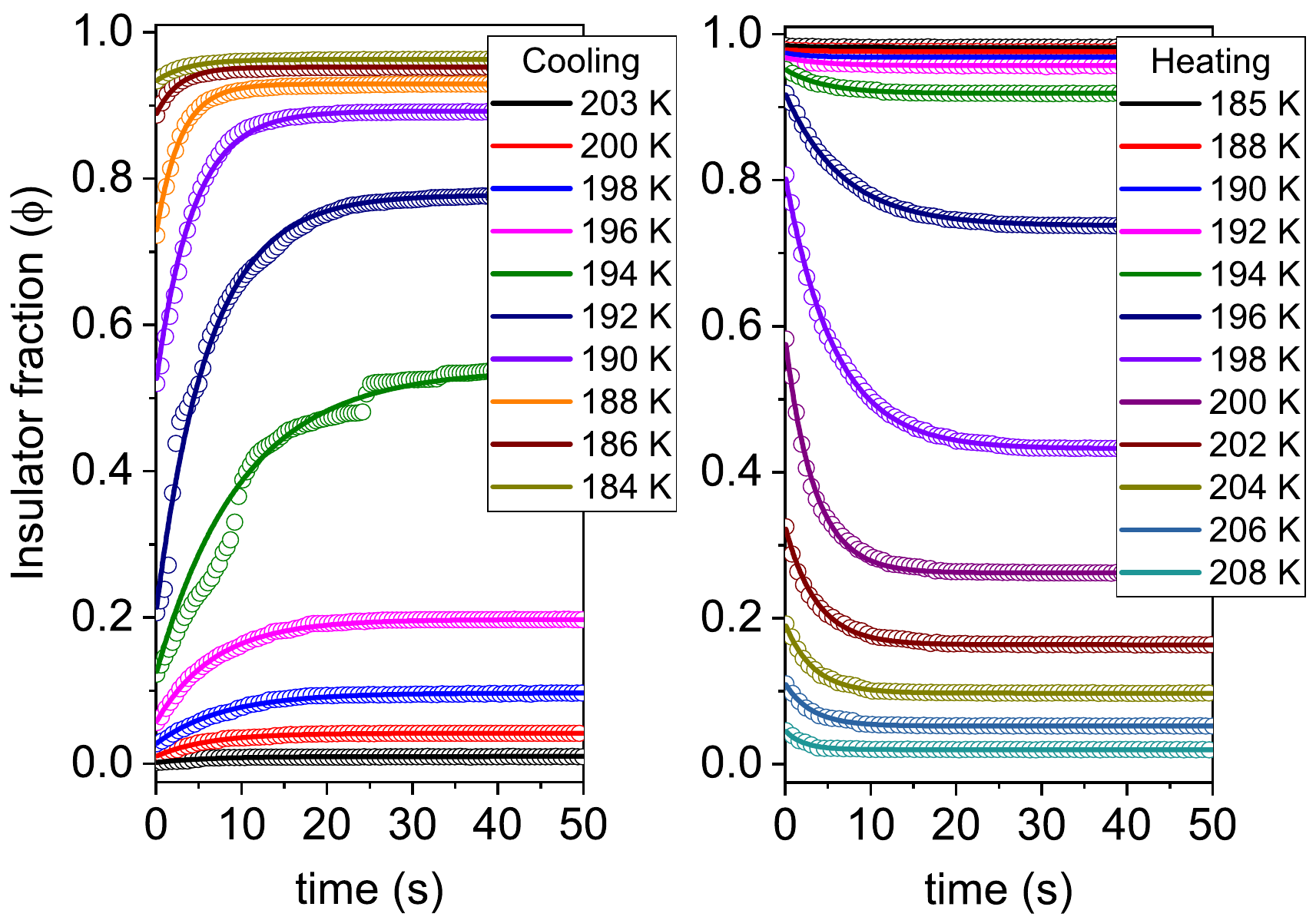}
\caption{Temporal evolution of order parameter (insulator fraction) during wait after shock heating (a) and cooling (b). Solid lines are exponential fits.}
\label{QAW_fit}
\center
\end{figure}

In Fig. \ref{quench_wait}, we have plotted the insulator fraction or order parameter ($\phi$) as a function of time and temperature. The quench-and-hold measurements' temperature starts from a perfectly insulating state (or metallic state). The sample is heated up (or cool down) rapidly to a specific set temperature, then allowed to equilibrate. Simultaneously, the sample's resistance was recorded and converted to the insulator fraction (order parameter $\phi$) using a percolation model based on McLachlan's general effective medium theory \cite{Insulator_frac}. The model reads
\begin{equation}\label{GEM_eq}
\phi\frac{(\sigma_I^{1/t} - \sigma_E^{1/t})}{(\sigma_I^{1/t} + A\sigma_E^{1/t})} + (1-\phi)\frac{(\sigma_M^{1/t} - \sigma_E^{1/t})}{(\sigma_M^{1/t} + A\sigma_E^{1/t})} = 0,
\end{equation}
where $\sigma_M$ and $\sigma_I$ are the conductivities of metallic and insulating phases. $A$ is defined as $(1-\phi_c)/\phi_c$ where $\phi_c$ is the insulator fraction at the percolation threshold. $\phi_c = 0.16$ and the critical exponent $t = 2$ for three dimensional system.

Figure \ref{QAW_fit} shows the first 50 seconds of phase evolution after reaching the set temperature, and the phase ordering relaxation times $\tau$ are inferred by the fitting equation given below
\begin{equation}\label{exp_eq}
\phi(t) = [\phi_{t=0}-\phi_{eq}]\exp{(-t/\tau)} + \phi_{eq}
\end{equation}
where $\phi_{eq}$ stand for quasi-static value of the order parameter.

\section{Dynamic Hysteresis in DTA measurements}
\label{App:DTA}
\begin{figure}[!h]
\center
\includegraphics[scale=0.3]{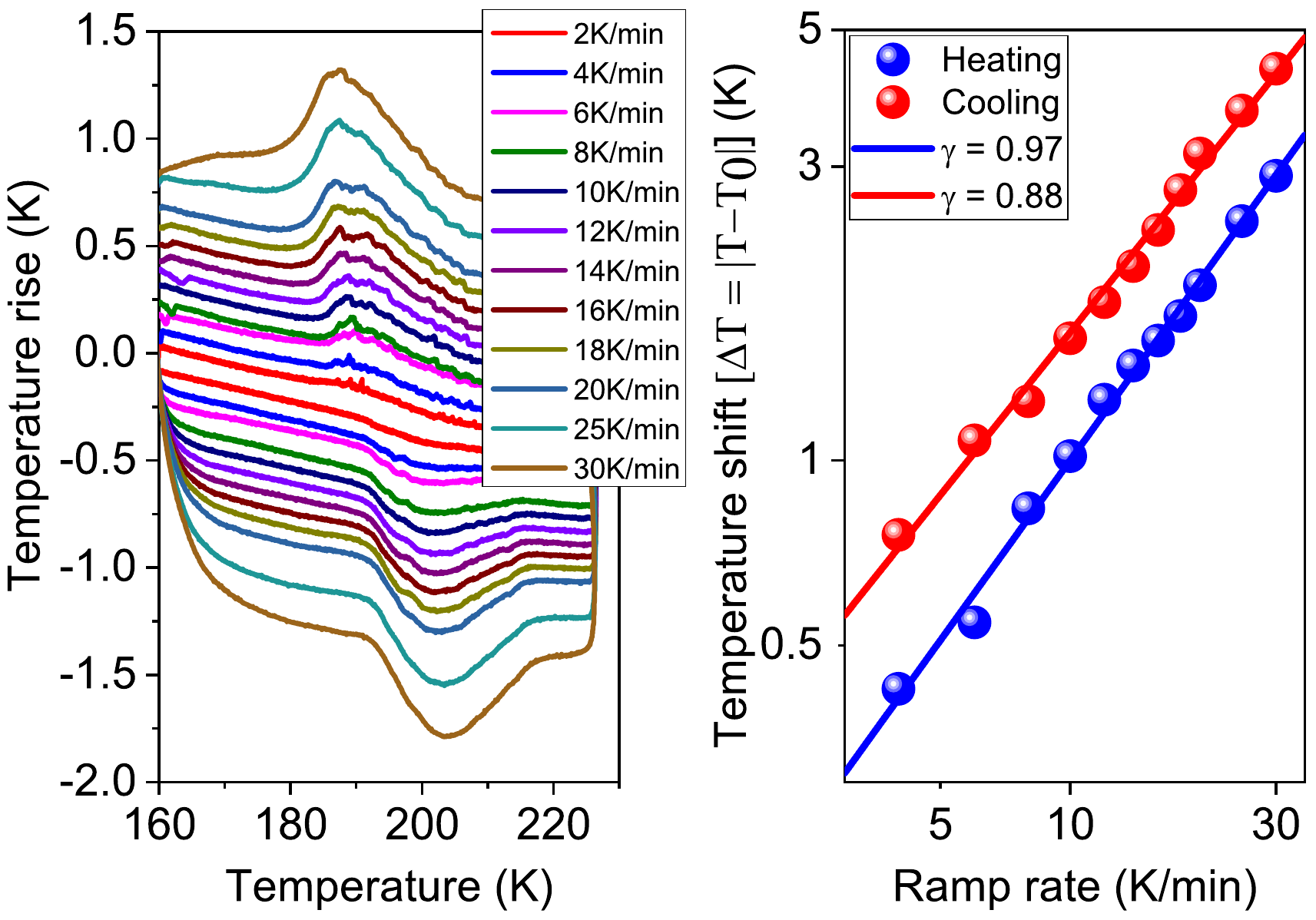}
\caption{(a) DTA signal as a function of temperature for different ramp rates. (c) The shift in transition temperatures flow scaling with temperature scanning rates through exponents $\Upsilon = 0.88$ (cooling) and $\Upsilon = 0.97$ (heating).}
\label{DTA}
\center
\end{figure}

Figure \ref{DTA} shows that the dynamic hysteresis scaling which has been observed in an independent thermodynamic measurement using the DTA technique. There is ambiguity in figuring out the transition temperatures from the DTA curve of a lower rate due to the multiple minor peaks. It is worth noting that the lower value of the data dominates the straight-line fits in a log-log graph. Therefore we narrowly consider the smoothed (degree 100) data with ramp 4 K/min and above to achieve scaling exponent from DTA measurements.
Nevertheless, the exponent values remain consistent with the preceding values extracted from resistance data. The transition temperatures, hence the exponent values, differ a bit depending upon the degrees of smoothness. That is why we may regard these measurements as rough. Note that the dynamical exponent in the heating branch is higher than the cooling branch of hysteresis. Despite having some limitations of the technique, the qualitative findings are somewhat supreme as the results are obtained from the thermodynamics measurement. 

\section{Zero temperature random field Ising mdoel}
\label{App:ZTRFIM}
This section provides the details on the algorithm for simulating the zero temperature random field Ising model (ZTRFIM). The Hamiltonian for the random field Ising model can be written as
\begin{equation}
\mathcal{H} = -J \sum_{\langle i,j \rangle} s_i s_j - \sum_{i} \left[ H(t) + h_i \right] s_i,
\label{eqn:RFIM}
\end{equation}
The Ising-like spins, $s_i=\pm 1$ are on three-dimensional cubic lattice of linear dimension $ L $ with periodic boundary conditions. The nearest neighbor spins interact ferromagnetically with interaction strength denoted by $ J $. We set $J=1$ for this work. All the spins on the lattice experience an external time-dependent uniform field, $H(t)$. The form of $H(t)$ depends on the real situation we would like to simulate and is discussed in detail below. Additionally, a random site-dependent but static magnetic field, $h_i$ is also present. The random local $h_i$ is chosen from a random Gaussian distribution, $V(h)$ given by
\begin{equation}
V(h) = \frac{1}{\sqrt{2 \pi \sigma^2}} e^{-h^2/(2\sigma^2)},
\label{eqn:Gaussian}
\end{equation}
where $ \sigma $ is the width of the distribution and denotes the disorder strength in this model. We average all the physical quantities using 50-100 independent disorder realizations. Here the random field Ising model is simulated at the zero-temperature, and therefore, there are no thermal fluctuations. The spin-flip is entirely determined by the sign change of the local field
\begin{equation}
E_i=J\sum_j s_j+ h_i+H, 
\label{eqn:flip}
\end{equation}
where the sum over $j$ is over the neighbor sites of $i$. For a cubic lattice we have six neighbors at each site.  This local field determines the spin flip completely at each time step as there is no thermal fluctuations.

\subsection{Dynamic hysteresis}
\label{App:ZTRFIM_rate}
We use a linear ramping of the field, ${H(t)=H_0 + R t}$ where $R$ is the rate of change of the magnetic field, and $H_0$ is the initial external field strength at $t=0$. The algorithm to obtain the magnetization at each step as the external field is linearly increased is presented below.
\begin{enumerate}
\item We set the spin for every site to $s_i=-1$. Similarly, we fix the field to $H(0)=-H_{0}$, where at $-H_0$, all the spins are expected to be in the down configuration, i.e., we have a fully polarized system.
\item We increase the external magnetic field by $R$ such that $H(t)=H(t-1)+R$.
\item We check all the sites if the local field defined in Eq.~(\ref{eqn:flip}) changes sign on any over the lattice.
\item We flip the spins of the sites where the local field $E_i$ changes sign.
\item Next, we move to the next time step by increasing the field by $R$. Hence, we repeat steps (2-5) until all the spins are flipped.
\end{enumerate}

We highlight the difference between this dynamic algorithm and the quasi-static simulations of ZTRFIM in Ref.~\cite{Sethna_prl93,Sethna_prl95}. For a quasi-static process that corresponds to $R \rightarrow 0$ in our notation, the system is allowed to equilibrate before increasing the magnetic field. To perform the equilibration, if a spin flips in step (3), the local energy of all sites needs to be re-evaluated to confirm whether the initial flip triggers the flipping of other spins. This procedure continues until none of the local energy flips sign for a set magnetic field $H(t)$. Only after ensuring that we have reached equilibrium for that $H(t)$, the magnetic field is increased further. Here, the equilibration procedure is bypassed in the linear ramping to capture the non-equilibrium nature of the physical quantities. 

\begin{figure}[!h]
\center
\includegraphics[scale=0.3]{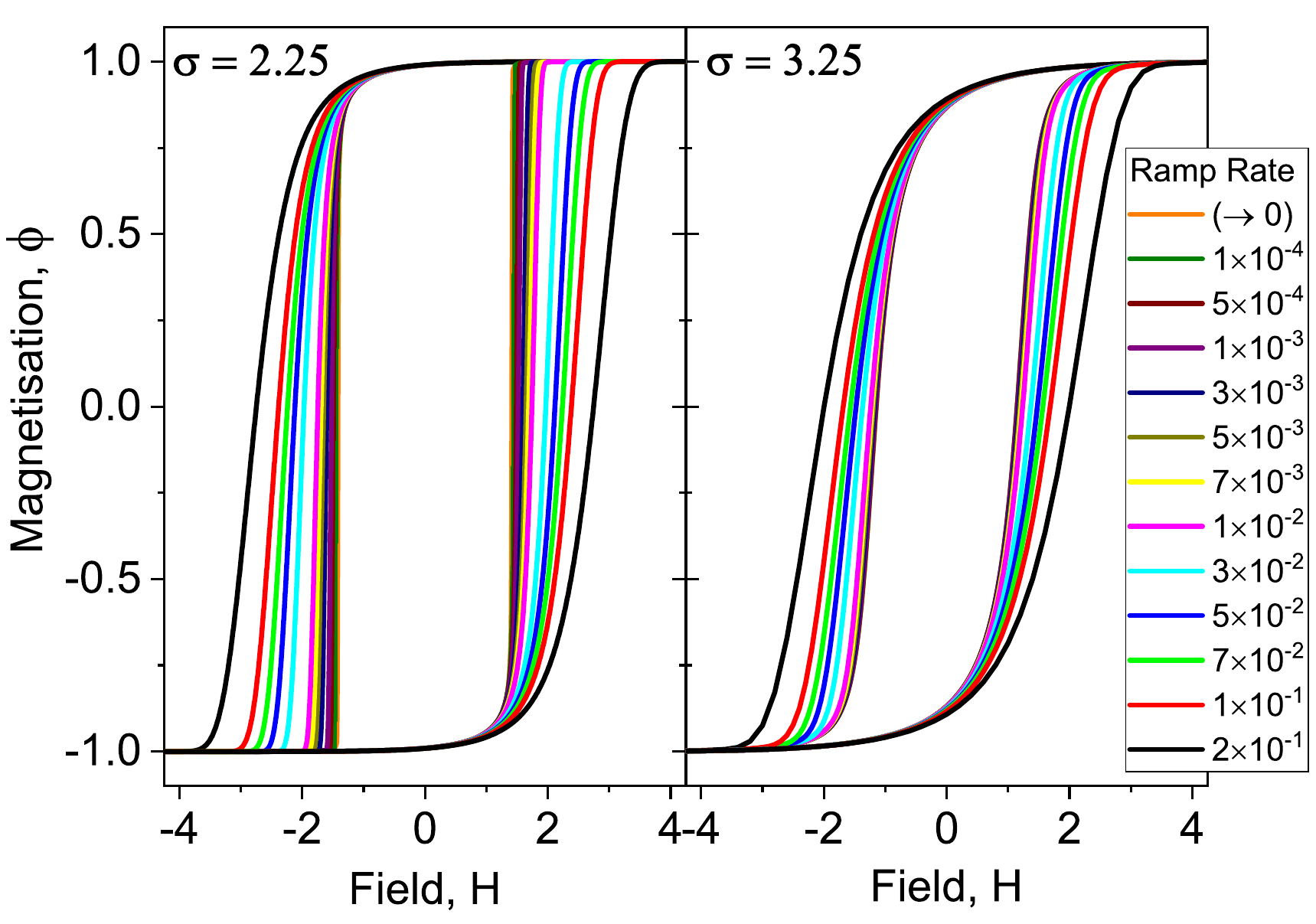}
\caption{Magnetization $\phi$ (order parameter) versus external field H for different ramp rate yield in the 3D-ZTRFIM simulations of system size $256^3$ under periodic boundary conditions  for disorder strength $\sigma$ = 2.25, and 3.25. Ramp rate ($\rightarrow$ 0) indicate the steady state hysteresis}
\label{Hysteresis}
\center
\end{figure}

The algorithm presented above can be extended when all the spins are in up-configuration, as we linearly decrease the field from $H(0)=H_{0}$ with the rate $R$. Therefore, the increasing and decreasing magnetic field leads to the two branches of the magnetization-field graph in Fig. \ref{Hysteresis}.

In Fig. \ref{Hysteresis}, we show that for a prolonged ramping, the magnetization curve indeed approaches the quasi-static value. Note that hysteresis is symmetric to the origin [Fig. \ref{Hysteresis}]. The shift in coercive fields $H_c(R)$, the fields at which the magnetization changes the sign, from the steady-state coercive field $H_c(R=0)$ fulfill a scaling relation with fields rate R:  
\begin{equation}
\Delta H_c (R)=|H_c(R)-H_c(0)|\propto R^\Upsilon.
\end{equation}

\begin{figure}[!h]
\center
\includegraphics[scale=0.32]{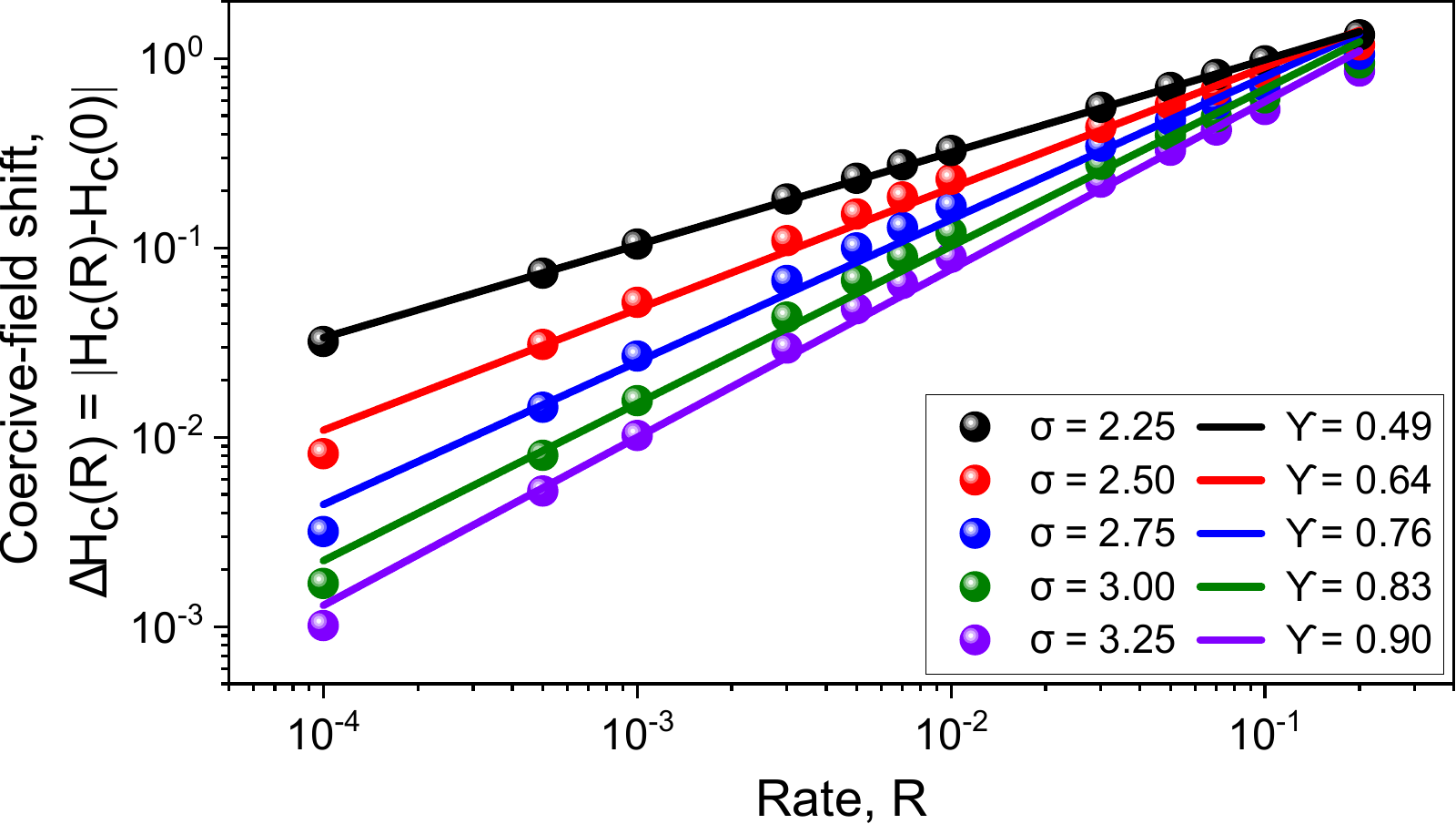}
\caption{The shift in coercive fields $H_c(R)$ from the steady-state coercive fields $H_c(0)$ follow a power law with ramp rate. (---) represent linear fitting with exponent $\Upsilon$.}
\label{Scal_Fitting}
\center
\end{figure}

Figure \ref{Scal_Fitting} shows the power-law fitting for the different disorders, and the fitting deteriorates with the increasing disorder from the critical disorder. The scaling exponent versus disorder graph has been shown in Fig. \ref{Dynamic_hysteresis}. It is worth noting that when the disorder strength higher than a threshold value $\sigma_{th} \approx 3.30$, the fitting with a single exponent will not be feasible (see Appendix \ref{App:Error}).  
  
\subsubsection*{Zero temperature mean field}
Furthermore, we also performed a zero temperature mean-field analysis of same model. In order to do so, the spin flip is determined by,
\begin{equation}
E_i=J z m+ h_i+H, 
\label{eqn:MFflip}
\end{equation}
where $z$ is the average number of nearest neighbors and $m$ is the average magnetization of the system $m=\sum_{i=1}^{N} s_i$. Note that here the local field is determined by the average magnetization of the whole system instead of neighboring spin. Using the spin-flip protocol of equation \ref{eqn:MFflip} we perform the same algorithm presented in the previous section.

\subsection{Time constant}
\label{App:ZTRFIM_time}
We also find the time required for a fully polarized system to relax to a steady state when the external field is suddenly quenched to a certain value, mimicking the quench-and-hold experiment. The algorithm for the same is presented below:
\begin{enumerate}
\item First we set the spins on every site to $s_i=-1$ or $s=1$. 
\item Next, we fix the external magnetic field to $H=H_f$ for which we need to find the time to reach steady-state. We also initialize the time constant to $\tau=0$.
\item We check all the sites if the local field defined in Eq.~(\ref{eqn:flip}) changes sign on any of the sites.
\item If there is a sign change of the local field on any site, we flip the spin on that site.
\item If in step 4 there is at least a single flip over the whole lattice, we increase the unit of time $\tau=\tau+1$ and then again repeat the steps from (3-5).
\item If during the check, in step 3, there is no flip over the whole lattice, we exit the process. The time during the exit is the relaxation time $\tau$.
\end{enumerate}

\begin{figure}[!h]
\center
\includegraphics[scale=0.28]{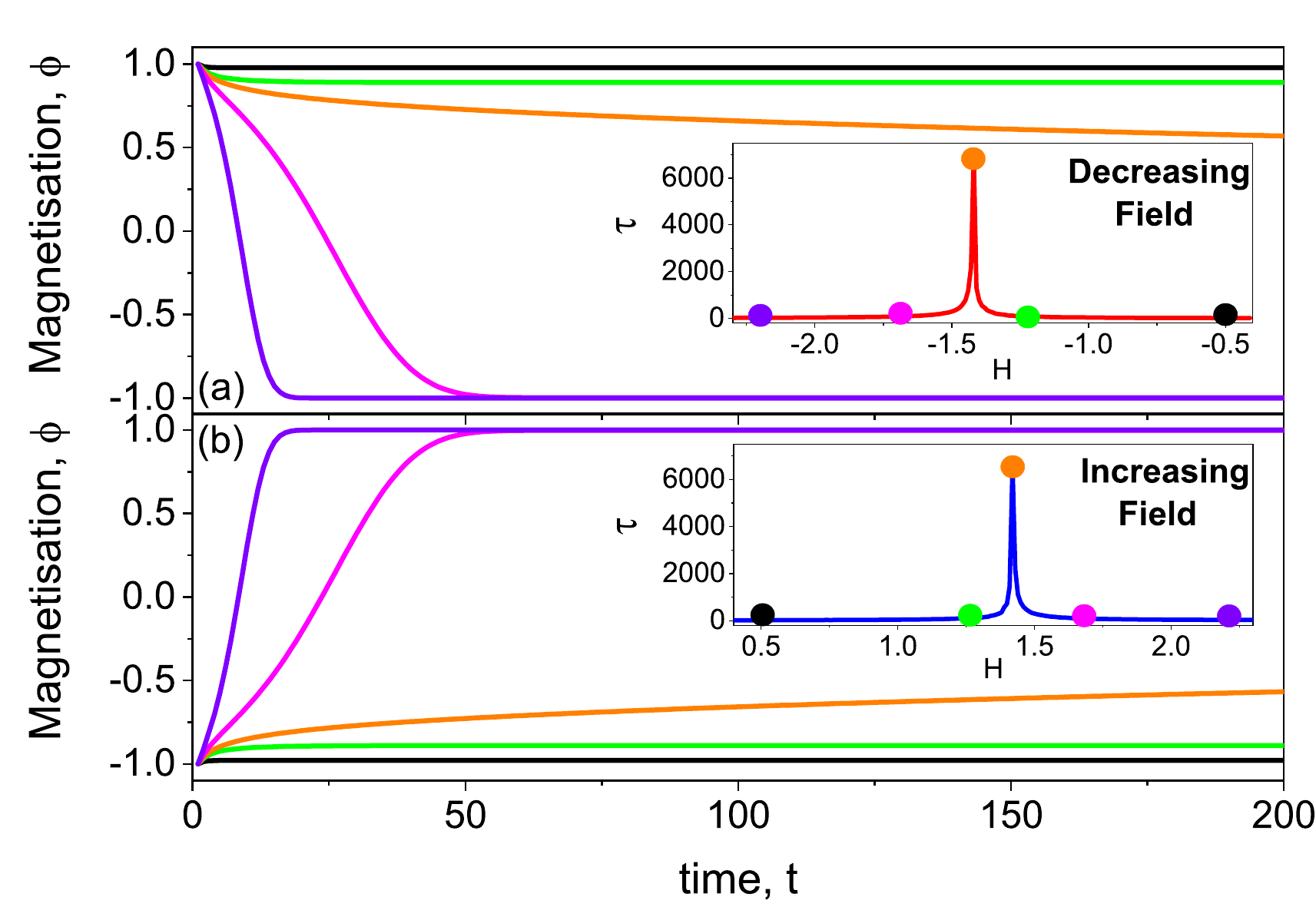}
\caption{Temporal evolution of order parameter $\phi$ in phase-ordering simulation on ZTRFIM of system size $256^3$ for $\sigma = 2.25$. The individual color represents the distinct quenched-field shown in the equilibration time (relaxation time) vs. quenched-field graph (inset) for decreasing (a) and increasing (b) the fields.}
\label{Evolution}
\center
\end{figure}

The relaxation time results are presented in Fig. \ref{eq_time}(a) and \ref{eq_time}(b). This time-scale arises because the system relaxes to a steady state by series of small avalanches~\cite{Sethna_prl93,Sethna_prl95}. Figure \ref{Evolution} shows the time evolution of the magnetization when the field is quenched to a particular value, and the magnetization $\phi$ finally reached the steady-state after time $\tau$. Equilibration time $\tau$ peaks at the transition points (coercive fields) [Fig. \ref{eq_time} and Fig. \ref{Evolution}(inset)], demonstrates that dynamics slow down considerably near the spinodal point.

\section{Self-averaging property and disorder-induced error}
\label{App:Self_averaging}
In this section, we study the self-averaging property of the random field Ising model system that concerns the dependence of physical quantity, let us say the magnetization $\phi(H)$, with the disorder. If $\phi(H)$ is a self-averaging quantity, most disorder realizations will provide the same value of $\phi(H)$ in the thermodynamic limit. In that case, a very few numbers of sample average is good enough to provide an appropriate $\phi(H)$ through numerical simulation. If the system is not self-averaging, the magnetization strongly depends on the disorder realization. Even in an infinite system, one would not get a meaningful result from a single realization; demands for repetitive measurements over many samples \cite{Vives_prb06}. 

The self-averageness of the RFIM system of disorder strength $\sigma$ can be estimated from the relative variance of magnetization, 
\begin{equation}
R_m = \frac{[\langle m(H)^2 \rangle -\langle m(H) \rangle^2]}{\langle m(H)\rangle^2},
\end{equation}
where $\langle ...\rangle$ denotes the average over disorder. The system is self-averaging, if $R_m \to 0$, and non self-averaging, if $R_m \propto const \neq 0$, as $L \to \infty$. Whereas, the system is consider to be weak self-averaging if $R_m \propto L^{(\alpha/\nu)}$ for $L \to \infty$ \cite{Fytas_pre16}.

\begin{figure}[!h]
\center
\includegraphics[scale=0.29]{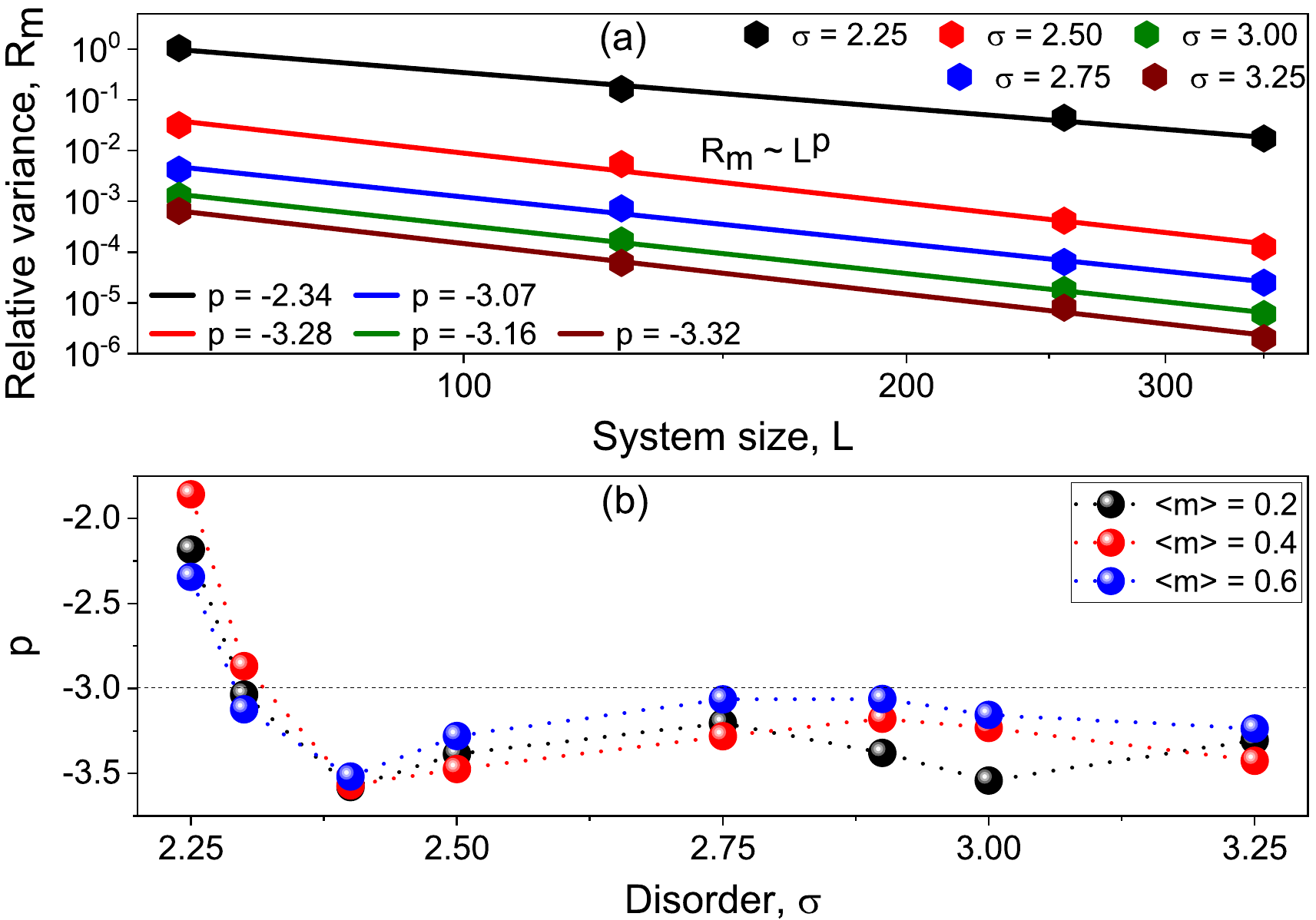}
\caption{(a) Relative variance $R_m$ of $m(H)$ calculated at $\langle m\rangle = 0.6$ of the $\phi$ vs. $H$ curve as a function of system size $L$ for different values of disorder $\sigma$. The $R_m$ follow a power law decay with the $L$, $R_m \sim L^p$. (b) The scaling factor $p$ calculated for selected values of $\langle m\rangle$ as a function of $\sigma$. For $\sigma > 2.30$, $R_m$ strongly decaying with the system size ($p > 3$). On the other hand, the decay is comparatively slow for $\sigma \leq 2.30$.}
\label{Selfavg}
\center
\end{figure}

Figure \ref{Selfavg} shows the system is always self-averaging for $\sigma > \sigma_c$ \cite{Vives_prb06}. However, we find that there are two regimes of self-averaging, strongly self-averaging for $\sigma > 2.3$ and weakly self-averaging for $\sigma \leq 2.30$. The weak self-averaging has been observed, as expected, near the pure critical point (at $\sigma = \sigma_c$), which is very different from the spinodal points. In the weak regime ($2.3 > \sigma > \sigma_c$), We have checked that the dynamical hysteresis exponent $\Upsilon$ converges within 10-20 disorder sample average. Although, the fitting error is much larger than the sampling error. However, the maximum disorder-induced errors can be estimated from the standard deviation of magnetization $\delta \phi$. We generate two set of hysteresis curve by adding and subtracting $\delta \phi(H)$ with the $\phi(H)$; and simultaneously calculated the two exponents $\Upsilon_{\phi + \delta \phi}$ and $\Upsilon_{\phi - \delta \phi}$. The different between the two exponents $\delta \Upsilon _{disorder} = |\Upsilon_{\phi + \delta \phi} - \Upsilon_{\phi - \delta \phi}|$ can be consider as maximum disorder-induced error. In the strong self-averaging region ($\sigma > 2.3$), the disorder-induced error $\delta \Upsilon _{disorder}$ is an order of magnitude lower than the fitting error $\delta \Upsilon _{fitting}$. Whereas in the weakly self-averaged regime, the disorder error is larger than the fitting error. The total errors, the sum of fitting and disorder-induced errors, are exhibit in Fig. \ref{Dynamic_hysteresis}. The fitting error has been described in the next section.

\section{Power law fitting and Error}
\label{App:Error}
The dynamical renormalized shifts $\Delta T\ (|T^{i}_0-T^{i}_{obs}(R)|,\ i= heat\ or\ cool)$ follows scaling with temperature rates R such as
\begin{equation}\label{Scaling}
|T^{i}_0-T^{i}_{obs}(R)| = aR^{\Upsilon}.
\end{equation}
The exponent $\Upsilon$ was extracted by fitting the above equation [Eq. \ref{Scaling}] with the experimentally and numerically obtained data points. We analyzed the data set twice using the best straight-line fitting method and statistical distribution of the non-linear fitting method. 

{\it The best straight-line fitting:} In equation \ref{Scaling}, there are three unknown parameters, $T^{i}_0$ the quasistatic transition temperature, $\Upsilon$ and the constant a. The constant a could be bypassed by fitting a straight line in the log-log graph, where the slope of the straight line would be the $\Upsilon$. We varied the $T^{i}_0$ within the acceptable region to achieve the best straight line fit. The acceptable values of $T^{i}_0$ stand below (or above) $T^{i}_{R1}$, the observed transition temperature under the lowest heating (or cooling) rate. Since $\Delta T$ is a monotonic increasing function of R, the acceptable region (let say $\delta T$) is bounded by transition temperature difference under two low rates ($|T^{i}_{R1}-T^{i}_{R2}|$) whose difference is larger than the lowest rate (i.e., $R2-R1 > R1$). Precisely $(T_{R1}-\delta T) < T_0 < T_{R1}$ and $(T_{R1}+\delta T) > T_0 > T_{R1}$ are the allowed quasi-static temperatures region for heating and cooling.

\begin{figure}[!h]
\center
\includegraphics[scale=0.3]{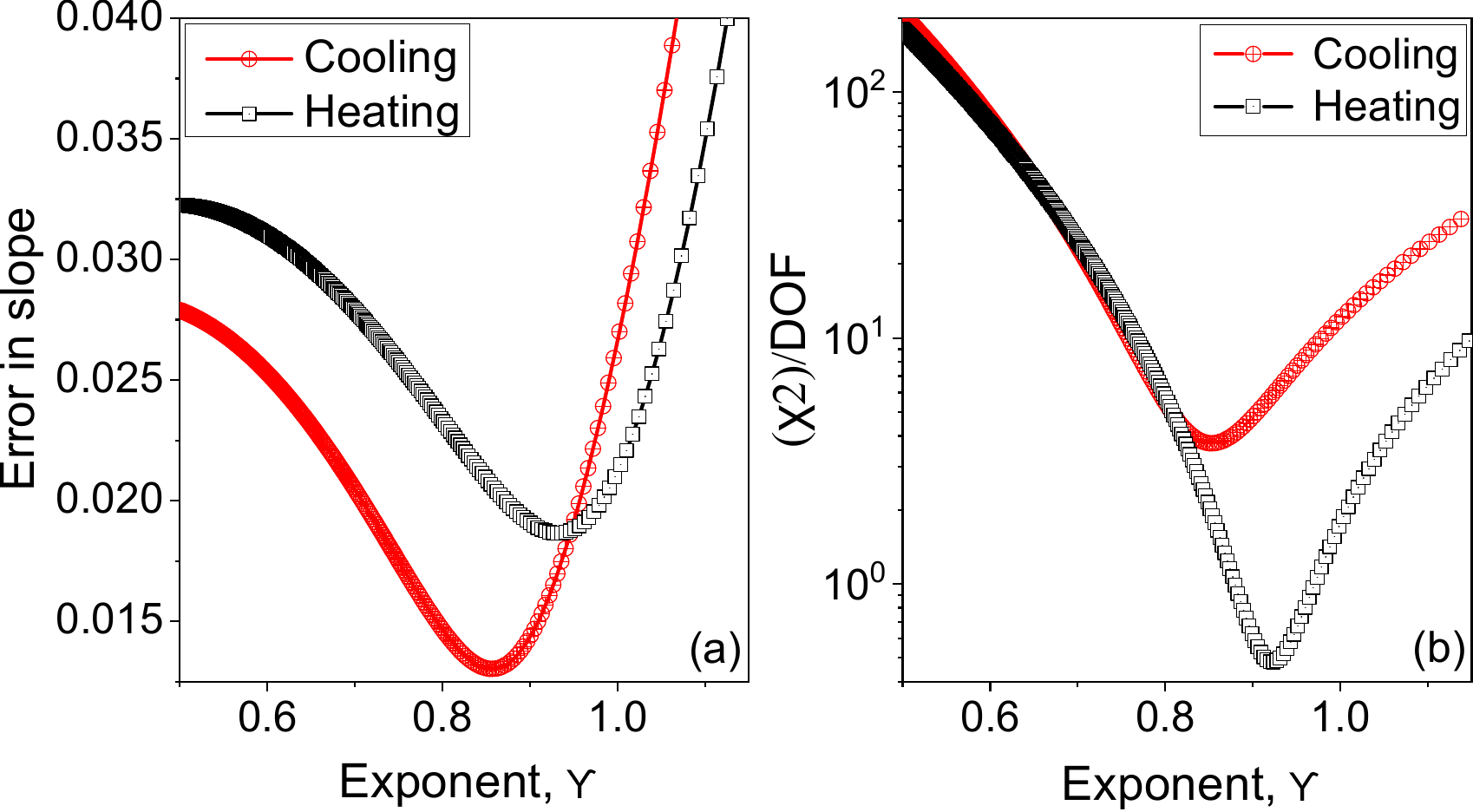}
\caption{(a) Least square fitting error vs the exponent $\Upsilon$ for different values of the quasi-static transition temperature $T_0$ and (b) corresponding reduced chi-square statistical quality $\chi^2/DOF$ vs $\Upsilon$. The minimum error are corresponding to minimum of $\chi^2/DOF$; $\Upsilon = 0.92$ for heating and $\Upsilon = 0.85$ for cooling.}
\label{Minimum_error}
\center
\end{figure}

Figure \ref{Minimum_error} (a) shows that the least square error in the slope of the straight line as a function of the exponent. Each point of the graph corresponds to a hand-picked quasistatic transition temperature $T^{i}_0$. The exponent corresponding to the minimum of reduced chi-square statistical quality ($\chi^2/DOF$) could be thought of conclusive exponent. The $\chi^2/DOF$ in Fig. \ref{Minimum_error}(b) define as
\begin{equation}
\frac{\chi^2}{DOF} = \frac{1}{DOF}\sum_i\frac{(O_i-E_i)^2}{\Omega_i^2}.
\end{equation}  
where $O_i$ is the i-th observed data and $E_i$ is the i-th expected data of the straight-line fit. The degree of freedom (DOF) equals to the number of observations minus the number of fitting parameters. The values of $\Omega$, the uncertainty of transition temperatures, are not quite clear as it strongly depends on where we extract the transition temperature and how much incertitude is acceptable. In Fig. \ref{Minimum_error}(b) we calculate the $\chi^2/DOF$ by considering $\Omega = 0.003$ K to $0.1$ K for $R = 0.5$ K/min to $40$ K/min. The exponent is coming out to be $\Upsilon = 0.92$ for heating and $\Upsilon = 0.85$ for cooling, which is close to the exponents corresponding to the minimum of `error in slope' in Fig. \ref{Minimum_error}(a). 

However, the lower values of the data dominate in the straight line fitting on the log-log graph. Small changes in $T^{i}_0$, worsen the lower-rate transition temperature shift $\Delta T$, lead to a significant change in the exponent value. For these reasons, we come up with more logical data fitting kits where every data point contributes equally to extract the scaling exponent.

{\it Statistical distributions of non-linear fitting:} The simplest technique is to pick out any four data points out of the whole set of data and calculate the scaling exponent for all possibilities. The exponent's acceptance can be judged according to the quasi-static temperature, which has been calculated using that exponent. The histogram of all accepted exponents will follow a normal distribution where the mean value can be considered the final scaling exponent.

\begin{figure}[!h]
\center
\includegraphics[scale=0.3]{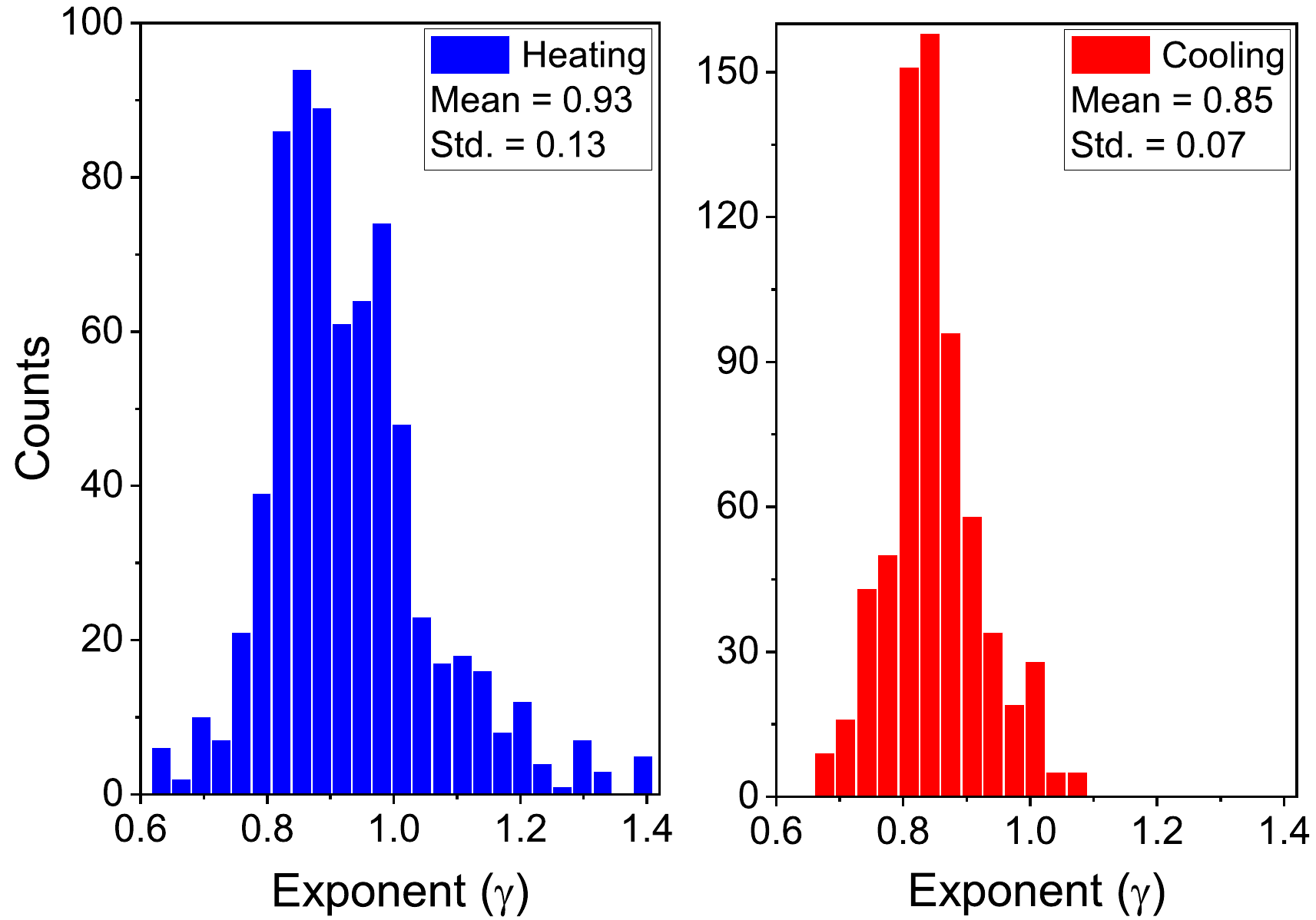}
\caption{Distribution of fitting exponent $\Upsilon$. The exponents were calculated using four independent points out of the whole data set, and every data point has an equal impact for extracting the fitting exponent.}
\label{Error_expt}
\center
\end{figure}

Let us consider $T_i$ and $T_j$ be the observed transition temperatures under ramp rate $R_i$ and $R_j$. We assume that shift in transition temperature from the quasi-static limits $T_0$ under temperature scanning rate will follow a power law:
\begin{equation} \label{error1}
T_{i} = T_0 + aR_{i}^{\Upsilon}, \hspace{10mm} T_{j} = T_0 + aR_{j}^{\Upsilon}  
\end{equation}
where the unknown constant a could be negative or positive depending upon the cooling or heating. 
The impact of quasi-static temperature to determine the exponent can be taken off by subtracting the above equations,
\begin{equation} \label{error2}
(T_i - T_j) = a(R_{i}^{\Upsilon} - R_{j}^{\Upsilon}).
\end{equation}
If N be the total total number of data points we can pick up two points in a $^NC_2$ (say $N1$) possible ways. We eliminate the unknown constant a by divide the equation \ref{error2} with the same equation for another pair of data \{k, l\}, i.e.
\begin{equation} \label{error3}
\frac{(T_i - T_j)}{(T_k - T_l)} = \frac{(R_{i}^{\Upsilon} - R_{j}^{\Upsilon})}{(R_{k}^{\Upsilon} - R_{l}^{\Upsilon})}
\end{equation}
Iterative numerical solutions of the above equation gives us $^{N1}C_2$ numbers of $\Upsilon$. 

For each $\Upsilon$ there are two $T_0$ 

 \begin{equation}
T^{\{i,j\}}_0 = \frac{T_i - (\frac{R_i}{R_j})^\Upsilon T_j}{1-(\frac{R_i}{R_j})^\Upsilon},\ 
T^{\{k, l\}}_0 = \frac{T_k - (\frac{R_k}{R_l})^\Upsilon T_l}{1-(\frac{R_k}{R_l})^\Upsilon}
 \end{equation}
 There is no restriction imposed on the value of $T_0$ which is not acceptable for equation \ref{error1} itself. The values of $\Upsilon$ for which the inferred $T_0$ lying inside the acceptable regions [$(T_{R1}-\delta T) < T^{heat}_0 < T_{R1}$ and $(T_{R1}+\delta T) > T^{cool}_0 > T_{R1}$] are allowed to draw the statistical distribution of exponent.
 
\begin{figure}[!t]
\center
\includegraphics[scale=0.29]{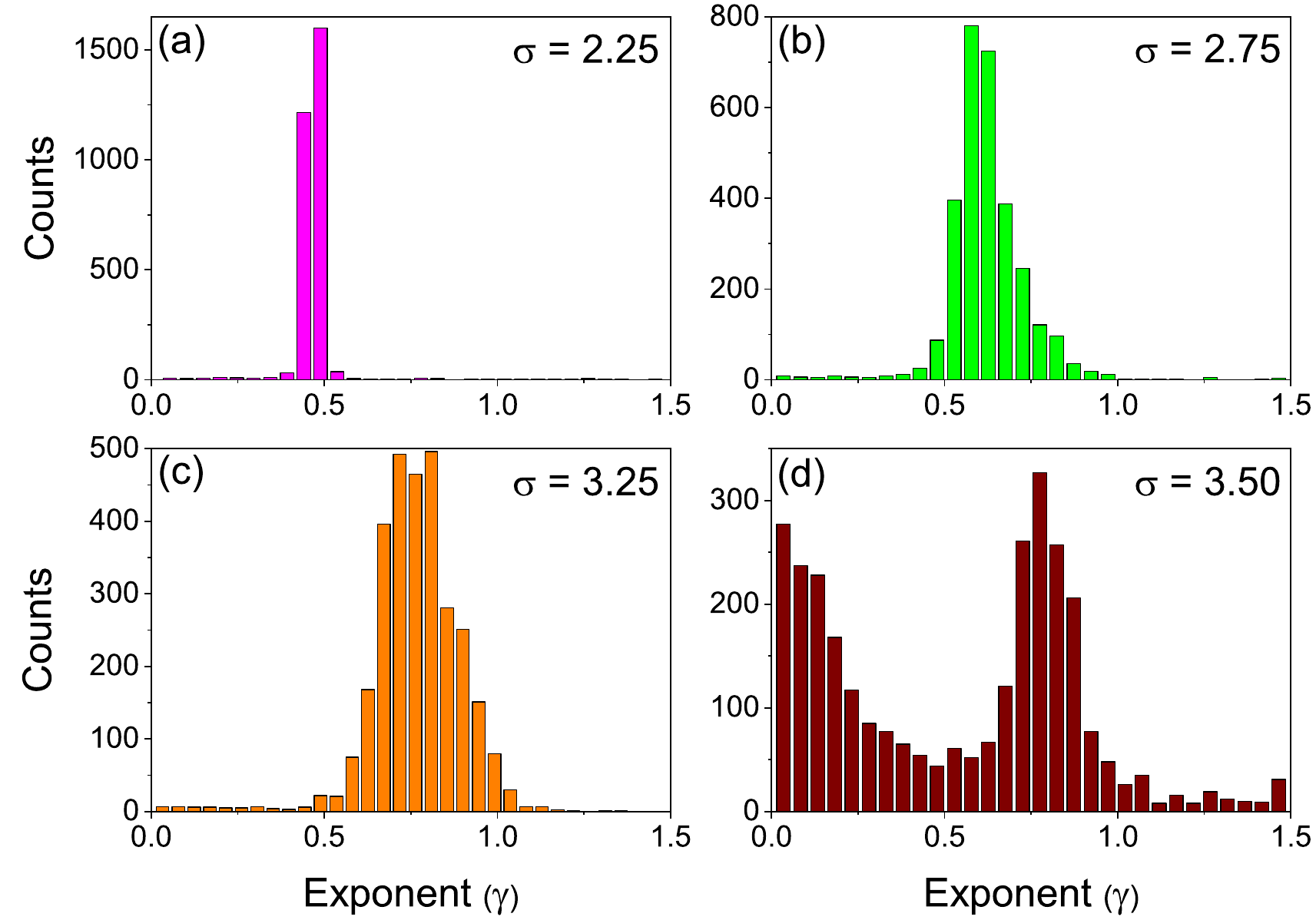}
\caption{Distribution of exponents for numerically evaluated dynamic hysteresis in ZTRFIM for disorder strength $\sigma$ = 2.25 (a), 2.75 (b), 3.25 (c), and 3.50 (d). The exponents ($\Upsilon$) are extracted from iterative numerical solutions of the equation \ref{error3}. The distribution broadened as the disorder strength increased (a to c) and finally above the threshold value ($\sigma_{th} \approx 3.30$) the distribution is no longer Gaussian (d).}
\label{Histo_dis}
\center
\end{figure}

The mean of the distribution [Fig. \ref{Error_expt}], $\Upsilon_{mean} = 0.93$ for heating and $\Upsilon_{mean} = 0.85$ for cooling, are compeer to the best straight-line fitting exponents [Fig. \ref{Minimum_error}]. The histogram's standard deviation, larger than the best straight line fitting error, can be considered the maximum error of the results. Note that the distribution in the cooling branch is quite sharp compared to the heating, and the error in heating (stander deviation) is twice the error in cooling.

{\bf Error in simulated exponent:} In the simulation, quasi-static temperature $T_0$ is known, equation \ref{Scaling} has only one unknown parameter in the log-log graph. The error in the straight line fitting can be used as estimates of error [Fig. \ref{Scal_Fitting}]. However, the goodness of the fitting is visible in the statistical distributions methods [Fig. \ref{Histo_dis}] although the mean of the distribution is lower than the straight-line fitting exponent. Note that the $\Upsilon$ are calculated using equation \ref{error3} without any restriction on the value of $T_0$. That causes some errors in the mean value.

\section{Free energy formalism of hysteresis}
\label{App:Free_energy}
A hysteresis arises because the system gets trapped in the metastable local minima and cannot reach the global minimum state within the experimental time scale. This is only possible when the system has insufficient fluctuations so that it can not cross the activation barriers separating the two phases. Depending upon the strength of fluctuations and barrier height, one can classified first-order phase transition (FOPT) into three classes.
\begin{figure}[!h]
\center
\includegraphics[scale=0.29]{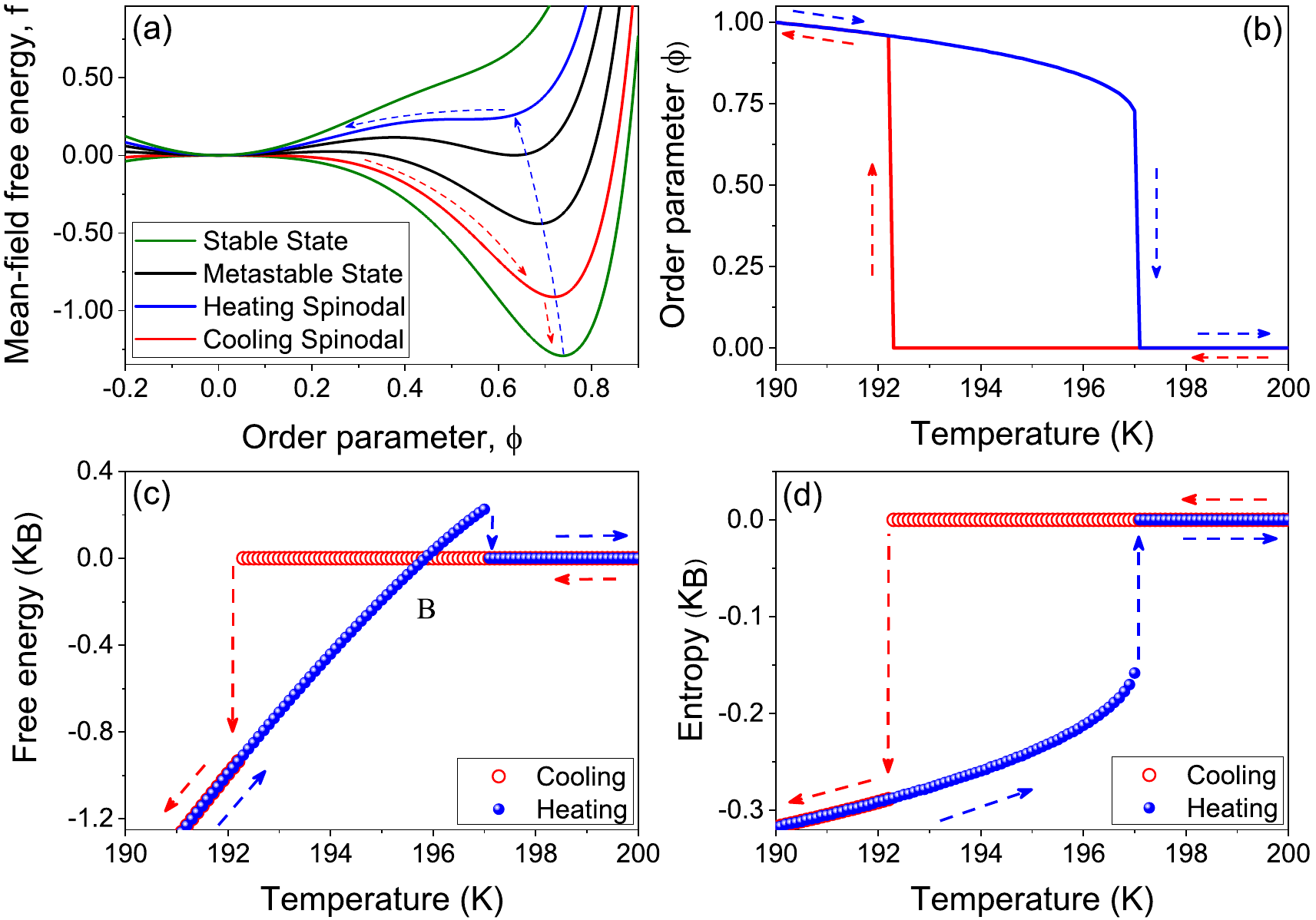}
\caption{(a) Graphical representation of free energy versus order parameter for different temperatures and associated (b) measurable order parameter versus temperature for an athermal system. The measurable order parameter values for different temperatures are computed from the (local or global) free energy minimum to which the system became trapped depending upon the previous history. The schematic diagram of free energy (c) and entropy (d) corresponding to the thermally driven hysteretic transition (b) are calculated from a  mean-field compressible Ising model \cite{Bar_prl18} where there are two free parameters, $T_c$ which represent the cooling transition point and other parameter $\xi$ fixed the heating transition points. We pick up $T_c = 192.28$ and $\xi = 0.1255$ just to fit experimentally obtained quasi-static transition temperatures. Arrows indicate the directions of heating (blue) and cooling (Red). It is important to note that the jump in free energy and entropy in thermally driven hysteretic transition is not restricted to the numerical model that we choose.}
\label{free_energy}
\center
\end{figure}

(i) When the fluctuations are significant, the system escapes from the local energy well to the global energy minimum through nucleation, and one would not observe hysteresis under equilibrium circumstances \cite{binder_rpp}. The phase transition takes place at the binodal point where the minimum of both the free energy well are in the same footing [3rd curve from the above in Fig.~\ref{free_energy}(a)]. This is the condition for the usual FOPT where the free energy is always convex under Maxwell's construction.

(ii) When the thermal fluctuations are insignificant in the kinetics of phase transformation (athermal), the metastable phase of a system can persist right up to the limit of metastability, spinodal point (mean-field concept), where the activation barrier against nucleation vanishes \cite{binder_rpp, chaikin-lubensky}. Here, the double-well free energy converted to a single well which is the usual case for continuous transitions [Blue and red curves in Fig.~\ref{free_energy}(a)]. Hence, one would expect to see the divergence of the correlation length along with divergence of the relaxation time of the order parameter as observed in a continuous transition. At those points of inflection, the free energy and entropy both are discontinuous, as observed in the schematic diagram Fig.~\ref{free_energy} (c), (d). In the experiment, the signature of such discontinuity is observed through hysteresis (free energy jump) accompanied by latent heat (entropy jump).

(iii) When the fluctuations are in-between the two extreme conditions mentioned above, one may observe the hysteresis along with latent heat but would not expect any signatures of criticality. In this case, the free energy and the entropy are also discontinuous, but jumps are comparatively less than the spinodal transitions [above and below the binodal point B in Fig.~\ref{free_energy}(c)]

In this article, we are in class (ii), where we experimentally observed dynamic scaling and critical slowing down (signature of continuous transition in the FOPT) differs this transition from the usual FOPT. This kind of athermal (or spinodal) transition arises due to the suppression of fluctuations by the long-range force during structural phase transition of many complex functional materials. \cite{Planes_book}. The concept of the spinodal first appeared in the van der Waals equation (1873), but it has been viewed only as an interesting artifact of the mean-field approximation \cite{binder_rpp, chaikin-lubensky, Bar_prl18}. However, we report that the disorder suppressed the spinodal instability and provided a non-mean-field dynamical exponent.


\begin{references}
\bibitem{binder_rpp} K. Binder, Theory of first-order phase transitions, Rep. Prog. Phys. {\bf 50}, 783 (1987).
\bibitem{Debenedetti} P. G. Debenedetti, {\it Metastable Liquids: Concepts and Principles} (Princeton University Press, 1996).
\bibitem{Planes_book} T. Kakeshita, T. Fukuda, A. Saxena, and A. Planes, {\it Disorder and Strain-Induced Complexity in Functional Materials} (Springer Berlin Heidelberg, 2012).
\bibitem{Planes_prl04} F. J. P\'erez-Reche, B. Tadic, L. Ma\~nosa, A. Planes, and E. Vives, Driving Rate Effects in Avalanche-Mediated First-Order Phase Transitions, Phys. Rev. Lett. {\bf 93}, 195701 (2004).
\bibitem{Liu_prl16} C. Liu, E. E. Ferrero, F. Puosi, Jean-Louis Barrat, and K. Martens, Driving Rate Dependence of Avalanche Statistics and Shapes at the Yielding Transition, Phys. Rev. Lett. {\bf 116}, 065501 (2016).
\bibitem{exchange_bias} S Giri, M Patra and S Majumdar, Exchange bias effect in alloys and compounds, Journal of Physics: Condensed Matter {\bf 23}, 073201 (2011).
\bibitem{kinetic_arrest}  W. Ito, K. Ito, R. Y. Umetsu, and R. Kainumaa, Kinetic arrest of martensitic transformation in the NiCoMnIn metamagnetic shape memory alloy, Appl. Phys. Lett. {\bf 92}, 021908 (2008).
\bibitem{Binder_pra84} K. Binder, Nucleation barriers, spinodals, and the Ginzburg criterion, Phys. Rev. A. {\bf 29}, 341 (1984).
\bibitem{Planes_prl01} F. J. P\'erez-Reche, E. Vives, L. Ma\~nosa, and A. Planes, Athermal Character of Structural Phase Transitions, Phys. Rev. Lett. {\bf 87}, 195701 (2001).
\bibitem{Zacharias_prl12} M. Zacharias, L. Bartosch, and M. Garst, Mott Metal-Insulator Transition on Compressible Lattices, Phys. Rev. Lett. {\bf 109}, 176401 (2012).
\bibitem{Rasmussen_prl01} K. \O{}. Rasmussen, T. Lookman, A. Saxena, A. R. Bishop, R. C. Albers, and S. R. Shenoy, Three-Dimensional Elastic Compatibility and Varieties of Twins in Martensites, Phys. Rev. Lett. {\bf 87}, 055704 (2001).
\bibitem{Bar_prl18} T. Bar, S. K. Choudhary, Md. A. Ashraf, K. S. Sujith, S. Puri, S. Raj, and B. Bansal, Kinetic Spinodal Instabilities in the Mott Transition in V$_2$O$_3$: Evidence from Hysteresis Scaling and Dissipative Phase Ordering, Phys. Rev. Lett. {\bf 121}, 045701 (2018).
\bibitem{Procaccia_pnas17} G. Parisi, I. Procaccia, C. Rainone, and M. Singh, Shear bands as manifestation of a criticality in yielding amorphous solids, Proc. Natl. Acad. Sci. USA {\bf 114}, 5577 (2017).
\bibitem{klein-monette} C. Unger, and W. Klein, Nucleation theory near the classical spinodal, Phys. Rev. B {\bf 29}, 2698 (1984) for a review, see L. Monette, Spinodal Nucleation, Int. J. Mod. Phys. B {\bf 8}, 1417 (1994).
\bibitem{Kundu_prl20} S. Kundu, T. Bar, R. K. Nayak, and B. Bansal, Critical slowing down at the abrupt mott transition: when the first-order phase transition becomes zeroth order and looks like second order, Phys. Rev. Lett. {\bf 124}, 095703 (2020).
\bibitem{Klein_pseudospinodal} N. Gulbahce, H. Gould, and W. Klein, Zeros of the partition function and pseudospinodals in long-range Ising models, Phys. Rev. E {\bf 69}, 036119 (2004).
\bibitem{Basov_nat18} K. W. Post, A. S. McLeod, M. Hepting, M. Bluschke, Y. Wang, G. Cristiani, G. Logvenov, A. Charnukha, G. X. Ni, P. Radhakrishnan, M. Minola, A. Pasupathy, A. V. Boris, E. Benckiser, K. A. Dahmen, E. W. Carlson, B. Keimer, and D. N. Basov , Coexisting first- and second-order electronic phase transitions in a correlated oxide, Nat. Phys. {\bf 14}, 1056 (2018).
\bibitem{Cao_prb90} W. Cao, J. A. Krumhansl, and Robert J. Gooding, Defect-induced heterogeneous transformations and thermal growth in athermal martensite, Phys. Rev. B {\bf 41}, 11319 (1990).
\bibitem{Jerky_DTA_peak} L. Z. T\'oth, S. Szab\'o, L. Dar\'oczi, and D. L. Beke, Calorimetric and acoustic emission study of martensitic transformation in single-crystalline ${\mathrm{Ni}}_{2}\mathrm{MnGa}$ alloys, Phys. Rev. B {\bf 90}, 224103 (2014).
\bibitem{Klein_pre07} H. Wang, H. Gould, and W. Klein, Homogeneous and heterogeneous nucleation of Lennard-Jones liquids, Phys. Rev. E {\bf 76}, 031604 (2007).
\bibitem{Sethna_prl93} J. P. Sethna, K. Dahmen, S. Kartha, J. A. Krumhansl, B. W. Roberts, and J. D. Shore, Hysteresis and hierarchies: Dynamics of disorder-driven first-order phase transformations, Phys. Rev. Lett. {\bf 70}, 3347 (1993).
\bibitem{Nandi_prl16} S. K. Nandi, G. Biroli, and G. Tarjus, Spinodals with Disorder: From Avalanches in Random Magnets to Glassy Dynamics, Phys. Rev. Lett. {\bf 116}, 145701 (2016).
\bibitem{Zapperi_Roy} S. Zapperi, P. Ray, H. E. Stanley, and A. Vespignani, First-order transition in the breakdown of disordered media, Phys. Rev. Lett., {\bf 78}, 1408 (1997). Avalanches in breakdown and fracture processes, Phys. Rev. E {\bf 59}, 5049 (1999).
\bibitem{Zheng_prb02} G. P. Zheng, and M. Li, Influence of impurities on dynamic hysteresis of magnetization reversal, Phys. Rev. B, {\bf 66}, 054406 (2002). While this paper considers the effect of quenched disorder in the finite temperature random-field Ising model, the average over the random field configurations is taken before the dynamical equation is solved (i.e., mean field estimation). Not surprisingly, they get $\Upsilon = 2/3$.
\bibitem{Ozawa_PNAS18} M. Ozawa, L. Berthier, G. Biroli, A. Rosso, and G. Tarjus, Random critical point separates brittle and ductile yielding transitions in amorphous materials, Proc. Natl. Acad. Sci. USA {\bf 115}, 6656 (2018).
\bibitem{earthquakes} D. S. Fisher, K. Dahmen, S. Ramanathan, Y. Ben-Zion, Statistics of earthquakes in simple models of heterogeneous faults, Phys. Rev. Lett. {\bf 78}, 4885 (1997). E. A. Jagla, F. P. Landes, A. Rosso, Viscoelastic effects in avalanche dynamics: A key to earthquake statistics, Phys. Rev. Lett. {\bf 112}, 174301 (2014).
\bibitem{QCD_prd20} M. A. Schindler, S. T. Schindler, L. Medina, and M. C. Ogilvie, Universality of pattern formation, Phys. Rev. D {\bf 102}, 114510 (2020).
\bibitem{social} J. P. Bouchaud, Crises and collective socio-economic phenomena: simple models and challenges, J. Stat. Phys. {\bf 151}, 567 (2013).
\bibitem{Bar_rsi21} T. Bar, and B. Bansal, Absolute calibration of the latent heat of transition using differential thermal analysis, Rev. Sci. Instrum. {\bf 92}, 075106 (2021).
\bibitem{Sethna_prl95} O. Perkovi\'c, K. Dahmen, and J. P. Sethna, Avalanches, Barkhausen Noise, and Plain Old Criticality, Phys. Rev. Lett. {\bf 75}, 4528 (1995).
\bibitem{Disorder_comment} However, at classical critical point, the type of extended disorder,  whether quench \cite{Perez-Reche_prl07, Cerrut_prb08, Vives_pre95} or anneal \cite{Perez-Reche_prb16}, is controversial for fine-tuning the criticality. Fortunately, we are apart from the classical critical point, and we also observed twining even after a fair amount of training the sample, and thus we would consider the quenched character of disorder for the modeling purposes.
\bibitem{Perez-Reche_prl07}F. J. Perez-Reche, L. Truskinovsky, and G. Zanzotto, Training-Induced Criticality in Martensites, Phys. Rev. Lett. {\bf 99}, 075501 (2007).
\bibitem{Cerrut_prb08} B. Cerrut, and E. Vives, Random-field Potts model with dipolarlike interactions: Hysteresis, avalanches, and microstructure, Phys. Rev. B {\bf 77}, 064114 (2008).
\bibitem{Vives_pre95} E. Vives, J. Goicoechea, J. Ort\'in, and A. Planes, Universality in models for disorder-induced phase transitions, Phys. Rev. E {\bf 52} R5 (1995).
\bibitem{Perez-Reche_prb16}F. J. Perez-Reche, C. Triguero, G. Zanzotto, and L. Truskinovsky, Origin of scale-free intermittency in structural first-order phase transitions, Phys. Rev. B {\bf 94}, 144102 (2016).
\bibitem{Fan_prb11} W. Fan, J. Cao, J. Seidel, Y. Gu, J. W. Yim, C. Barrett, K. M. Yu, J. Ji, R. Ramesh, L. Q. Chen, and J. Wu, Large kinetic asymmetry in the metal-insulator transition nucleated at localized and extended defects, Phys. Rev. B {\bf 83}, 235102 (2011).
\bibitem{Kang_ass14} L. Kang, L. Xie, Z. Chen, Y. Gao, X. Liu, Y. Yang, and W. Liang, Asymmetrically modulating the insulator-metal transition of thermochromic VO$_2$ films upon heating and cooling by mild surface-etching, Appl. Surf. Sci. {\bf 311}, 676 (2014).
\bibitem{Disorder_RPM} Return point memory has been observed above the critical disorder below that it is completely absent, see M. S. Pierce, C. R. Buechler, L. B. Sorensen, S. D. Kevan, E. A. Jagla, J. M. Deutsch, T. Mai, O. Narayan, J. E. Davies, K. Liu, G. T. Zimanyi, H. G. Katzgraber, O. Hellwig, E. E. Fullerton, P. Fischer, and J. B. Kortright, Disorder-induced magnetic memory: Experiments and theories, Phys. Rev. B {\bf 75}, 144406 (2007).
\bibitem{Catastrophe_theory} R. Gilmore, Catastrophe Theory for Scientists and Engineers (Dover, New York, 1981).
\bibitem{Analytic_continuation} X. An, D. Mesterh\'azy, and M. A. Stephanov, On spinodal points and Lee-Yang edge singularities, J. Stat. Mech. (2018) 033207.
\bibitem{Insulator_frac} D. Kumar, K. P. Rajeev, J. A. Alonso, and M. J. Mart\'inez-Lope, Journal of Physics: Condensed Matter {\bf 21}, 185402 (2009). K. H. Kim, M. Uehara, C. Hess, P. A. Sharma and S.-W. Cheong, Thermal and Electronic Transport Properties and Two-Phase Mixtures in La$_{5/8-x}$Pr$_x$Ca$_{3/8}$MnO$_3$, Phys. Rev. Lett. {\bf 84}, 2961 (2000).
\bibitem{athermal} H. Zheng, W. Wang, D. Wu, S. Xue, Q. Zhai, J. Frenzel, and Z. Luo, Athermal nature of the martensitic transformation in Heusler alloy Ni-Mn-Sn, Intermetallics, {\bf 36}, 90 (2013).
\bibitem{Basov_Nat17} A. S. McLeod, E. van Heumen, J. G. Ramirez, S. Wang, T. Saerbeck, S. Guenon, M. Goldflam, L. Anderegg, P. Kelly, A. Mueller, M. K. Liu, I. K. Schuller, and D. N. Basov, Nanotextured phase coexistence in the correlated insulator V$_2$O$_3$, Nat. Phys. {\bf 13}, 80 (2017).
\bibitem{Bagchi_prl07} P. Bhimalapuram, S. Chakrabarty, and B. Bagchi, Elucidating the Mechanism of Nucleation near the Gas-Liquid Spinodal, Phys. Rev. Lett. {\bf 98}, 206104 (2007).
\bibitem{Rao} M. Rao, H. R. Krishnamurthy, and R. Pandit, Magnetic hysteresis in two model spin systems, Phys. Rev. B {\bf 42}, 856 (1990). M. Rao and R. Pandit, Magnetic and thermal hysteresis in the O(N)-symmetric $(\Phi^2)^3$ model, Phys. Rev. B {\bf 43}, 3373 (1991). M. Rao, Comment on ``Scaling law for dynamical hysteresis'', Phys. Rev. Lett. {\bf 68}, 1436 (1992).
\bibitem{Fan_Zhong} F. Zhong, and J. Zhang, Renormalization Group Theory of Hysteresis, Phys. Rev. Lett. {\bf 75}, 2027 (1995). F. Zhong and Q. Chen, Theory of the Dynamics of First-Order Phase Transitions: Unstable Fixed Points, Exponents, and Dynamical Scaling, Phys. Rev. Lett. {\bf 95}, 175701 (2005). N. Liang and F. Zhong, Renormalization group theory for temperature-driven first-order phase transitions in scalar models, Front. Phys. {\bf 12}, 126403 (2017); F. Zhong, Renormalization-group theory of first-order phase transition dynamics in field-driven scalar model, Front. Phys. {\bf 12}, 126402 (2017).
\bibitem{Thermal_hysteresis} P. Jung, G. Gray, R. Roy, and P. Mandel, Scaling law for dynamical hysteresis, Phys. Rev. Lett. {\bf 65}, 1873 (1990). G. P. Zheng and J. X. Zhang, Thermal hysteresis scaling for first-order phase transitions, J. Phys.: Condens. Matter {\bf 10}, 275 (1998). W. Lee, J.-H. Kim, J. G. Hwang, H.-R. Noh and W. Jhe, Scaling of thermal hysteretic behavior in a parametrically modulated cold atomic system, Phys. Rev. E {\bf 94}, 032141 (2016).
\bibitem{glycerol} Y. Z. Wang, Y. Li, and J. X. Zhang, Scaling of the hysteresis in the glass transition of glycerol with the temperature scanning rate, J. Chem. Phys. {\bf 134}, 114510 (2011).
\bibitem{Shukla_pre18} P. Shukla, Hysteresis in the Ising model with Glauber dynamics, Phys. Rev. E {\bf 97}, 062127 (2018).
\bibitem{Pal_prb20} S. Pal, K. Kumar, and A. Banerjee, Universal scaling of charge-order melting in the magnetic field–pressure-temperature landscape, Phys. Rev. B {\bf 102}, 201109(R) (2020).
\bibitem{Imry_prb79} Y. Imry, and M. Wortis, Influence of quenched impurities on first-order phase transitions, Phys. Rev. B {\bf 19}, 3580 (1979).
\bibitem{Zhong_pre95} F. Zhong and J. X. Zhang, Scaling of thermal hysteresis with temperature scanning rate, Phys. Rev. E {\bf 51}(4),2898 (1995).
\bibitem{RRoy_prl95} A. Hohl, H. J. C. van der Linden, R. Roy, G. Goldsztein, F. Broner, and S. H. Strogatz, Scaling Laws for Dynamical Hysteresis in a Multidimensional Laser System, Phys. Rev. Lett. {\bf 74}, 2220 (1995).
\bibitem{short_range} The mean-field spinodal cannot in general survive in short-range low-dimensional (d = 2, 3) systems \cite{Zapperi_Roy} due to the existence of high enough thermal fluctuation but the long-range physics can be studied through RFIM at zero temperature \cite{Nandi_prl16} although the notion of spinodal instability has recently been observed in finite dimensional short-range interacting system \cite{Berthier_prl20} even at finite temperature \cite{Pelissetto_prl17}.
\bibitem{Berthier_prl20} L. Berthier, P. Charbonneau , and J. Kundu, Finite Dimensional Vestige of Spinodal Criticality above the Dynamical Glass Transition, Phys. Rev. Lett. {\bf 125}, 108001 (2020).
\bibitem{Pelissetto_prl17} A. Pelissetto and E. Vicari, Dynamic Off-Equilibrium Transition in Systems Slowly Driven across Thermal First-Order Phase Transitions, Phys. Rev. Lett. {\bf 118}, 030602 (2017).
\bibitem{Fytas_prl13} N. G. Fytas, and V. Mart\'{\i}n-Mayor, Universality in the Three-Dimensional Random-Field Ising Model, Phys. Rev. Lett. {\bf 110}, 227201 (2013).
\bibitem{Fytas_pre16} N. G. Fytas, and V. Mart\'{\i}n-Mayor, Efficient numerical methods for the random-field Ising model: Finite-size scaling, reweighting extrapolation, and computation of response functions, Phys. Rev. E {\bf 93}, 063308 (2016).
\bibitem{Footnote} The finite size effect will cancel out during steady-state subtraction and finally provide a robust dynamical exponent.
\bibitem{Moreira_prl12} A. A. Moreira, C. L. N. Oliveira, A. Hansen, N. A. M. Ara\'ujo, H. J. Herrmann, and J. S. Andrade, Jr., Fracturing Highly Disordered Materials, Phys. Rev. Lett. {\bf 109}, 255701 (2012).
\bibitem{Shekhawat_prl13} A. Shekhawat, S. Zapperi, and J. P. Sethna, From Damage Percolation to Crack Nucleation Through Finite Size Criticality, Phys. Rev. Lett. {\bf 110}, 185505 (2013).
\bibitem{Rizzo_prl13} T. Rizzo, Fate of the Hybrid Transition of Bootstrap Percolation in Physical Dimension, Phys. Rev. Lett. {\bf 122}, 108301 (2019).
\bibitem{Scheifele_pre13} B. Scheifele, I. Saika-Voivod, R. K. Bowles, and P. H. Poole, Heterogeneous nucleation in the low-barrier regime, Phys. Rev. E {\bf 87}, 042407 (2013).
\bibitem{Bhowmik_pre19} B. P. Bhowmik, S. Karmakar , I. Procaccia, and C. Rainone, Particle pinning suppresses spinodal criticality in the shear-banding instability, Phys. Rev. E {\bf 100}, 052110 (2019).
\bibitem{XRD} L. Ma, S. Q. Wang, Y. Z. Li, C. M. Zhen, D. L. Hou, W. H. Wang, J. L. Chen, and G. H. Wu, Martensitic and magnetic transformation in Mn$_{50}$Ni$_{50-x}$Sn$_x$ ferromagnetic shape memory alloys, J. Appl. Phys. {\bf 112}, 083902 (2012). Q. Tao, Z. D. Han, J. J. Wang, B. Qian, P. Zhang, X. F. Jiang, D. H. Wang, and Y. W. Du, Phase stability and magnetic-field-induced martensitic transformation in Mn-rich NiMnSn alloys, AIP Adv. {\bf 2}, 042181 (2012).
\bibitem{Ghosh_jap16} A. Ghosh, P. Sen, and K. Mandal, Measurement protocol dependent magnetocaloric properties in a Si-doped
Mn-rich Mn-Ni-Sn-Si off-stoichiometric Heusler alloy, J. Appl. Phys. {\bf 119}, 183902 (2016).
\bibitem{Ghosh_apl14} A. Ghosh and K. Mandal, Effect of structural disorder on the magnetocaloric properties of Ni-Mn-Sn alloy, Appl. Phys. Lett. {\bf 104}, 031905 (2014).
\bibitem{Ravel_apl02} B. Ravel, J. O. Cross, M. P. Raphael, V. G. Harris, R. Ramesh, and L. V. Saraf, Atomic disorder in Heusler Co$_2$MnGe measured by anomalous x-ray diffraction, Appl. Phys. Lett. {\bf 81}, 2812 (2002).
\bibitem{Avalanches_VO2} A. Sharoni, J. G. Ram\'irez, and I. K. Schuller, Multiple Avalanches across the Metal-Insulator Transition of Vanadium Oxide Nanoscaled Junctions, Phys. Rev. Lett. {\bf 101}, 026404 (2008).
\bibitem{Schuller_prb18} J. D. Valle, N. Ghazikhanian, Y. Kalcheim, J. Trastoy, M. H. Lee, M. J. Rozenberg, and I. K. Schuller, Resistive asymmetry due to spatial confinement in first-order phase transitions, Phys. Rev. B {\bf 98}, 045123 (2018).
\bibitem{Vives_prb06} X. Illa, M.-L. Rosinberg, and E. Vives, Influence of the driving mechanism on the response of systems with athermal dynamics: The example of the random-field Ising model, Phys. Rev. B {\bf 74}, 224403 (2006).
\bibitem{chaikin-lubensky} P. M. Chaikin, and T. C. Lubensky, {\it Principles of Condensed Matter Physics}, Cambridge University Press, Cambridge, UK (1995).
\end{references}
\end{document}